\documentclass[10pt,oneside]{article}
\usepackage{graphicx}
\usepackage{epstopdf}
\usepackage{amsmath}
\usepackage{latexsym}
\usepackage{amssymb}
\usepackage{amsfonts}
\usepackage{pifont}
\usepackage[small,bf,margin=0.5cm]{caption}
\usepackage{subfig}
\usepackage{tikz}
\usetikzlibrary{shapes,arrows}
\usetikzlibrary{calc}
\usepackage{pgf}
\usepackage{pgffor}
\usepackage{subfig}
\usepackage[english]{babel}
\usepackage{amsthm}

\usepackage[affil-it]{authblk}

\usepackage{hyperref}
\usepackage[all]{hypcap}    

\usepackage{lastpage}			
\usepackage{fullpage}			

\usepackage{blkarray}

\usepackage{fancyhdr}			
\usepackage[top=.75in,left=.9in,bottom=1.
in,right=.9in]{geometry}

\usepackage{hyperref}

\newcommand{\defi}{:=}

\newcommand{\trans}{^*}

\newcommand{\pinv}{^{+}}
\renewcommand{\Re}{\mathbb{R}}
\newcommand{\CC}{\mathbb{C}}

\newcommand{\normal}{\mathcal{N}}

\newcommand{\range}{\mathcal{R}}

\newcommand{\koop}{\mathcal{K}}
\newcommand{\statesp}{\mathcal{X}}

\newcommand{\admd}{A_\text{dmd}}

\newcommand{\proj}{\mathbb{P}}

\graphicspath{{figures_dynsys/}}

\begin{document}

\title{De-biasing the dynamic mode decomposition for applied Koopman spectral analysis of noisy datasets}


\author[1]{Maziar~S.~Hemati\thanks{Corresponding author: \href{mailto:mhemati@umn.edu}{mhemati@umn.edu}}}
\author[2]{Clarence~W.~Rowley}
\author[3]{Eric~A.~Deem}
\author[3]{Louis~N.~Cattafesta}
\affil[1]{{\small{Aerospace Engineering \& Mechanics, University of Minnesota, Minneapolis,~MN~55455}}}
\affil[2]{{\small{Mechanical and Aerospace Engineering, Princeton University, Princeton,~NJ~08544}}}
\affil[3]{{\small{Florida Center for Advanced Aero-Propulsion, Florida State University, Tallahassee,~FL~32310}}}

\date{}

\maketitle

\begin{abstract}
The Dynamic Mode Decomposition (DMD)---a popular method for performing data-driven Koopman 
spectral analysis---has gained increased adoption as a technique for extracting dynamically
meaningful spatio-temporal descriptions of fluid flows from snapshot measurements.
%
Often times, DMD descriptions can be used for predictive purposes as well, which enables informed 
decision-making based on DMD model-forecasts.
Despite its widespread use and utility, DMD regularly fails to yield accurate dynamical 
descriptions when the measured snapshot data are imprecise due to, e.g., sensor noise.
Here, we express DMD as a two-stage algorithm in order to isolate a source of systematic error.
We show that DMD's first stage, a subspace projection step, systematically introduces bias errors 
by processing snapshots asymmetrically.
To remove this systematic error, we propose utilizing an augmented snapshot matrix in a 
subspace projection step, as in problems of \emph{total least-squares}, in order to account for 
the error present in all snapshots.
The resulting unbiased and noise-aware \emph{total DMD} (TDMD) formulation reduces to standard DMD
in the absence of snapshot errors, while the two-stage perspective generalizes the de-biasing 
framework to other related methods as well.
TDMD's performance is demonstrated in numerical and experimental fluids examples.
\end{abstract}

\section{Introduction}
Dynamical systems---mathematical representations of a system's time-evolution---are of great 
importance and utility in the natural, social, and applied sciences, as they can provide a means 
of describing---and, therefore, better understanding---complex phenomena.
Often times, dynamical models can also be used in a predictive manner, to forecast the future 
behavior of a particular system, from which actionable decisions can be made.
Still, reliable and insightful models can be difficult to formulate in the context of nonlinear 
systems, such as fluid flows, which can exhibit complex behaviors on a broad range of spatial and 
temporal scales.
For instance, while fluid flows can be described mathematically from first-principles physics-based 
modeling (e.g.,~the Navier-Stokes equations), such models often lack closed-form solutions;
although numerical solutions can be sought, significant computational resources
may be demanded, which can make analysis and prediction unwieldy and impractical.
Even in instances for which a numerical solution can be reasonably computed, the resulting 
data---on their own---will not necessarily provide useful insight into the underlying 
characteristics of the fluid flow evolution.
Furthermore, for many dynamical systems, first-principles modeling can be
prohibitively challenging due to the sheer scale and complexity of the system dynamics; 
in such instances, the best recourse may be to determine a model from empirical data collected 
through system observations (i.e.,~a \emph{data-driven} approach).

In an effort to address these modeling challenges, the \emph{dynamic mode 
decomposition}~(\emph{DMD}) was developed in the fluid mechanics community as an equation-free 
data-driven technique capable of extracting dynamically relevant spatial structures and associated 
temporal characteristics (i.e.,~growth/decay rates and oscillation frequencies) from snapshot 
observations (e.g.,~pressure, velocity, vorticity) sampled from a fluid 
flow~\cite{schmidJFM2010,schmidAPS2008}.
It was shown in~\cite{rowleyJFM2009} that DMD approximates the \emph{Koopman 
operator}~\cite{koopmanPNAS1931}, an infinite-dimensional {\em linear} operator that describes the 
evolution of a {\em nonlinear} dynamical system by its action on observables (defined precisely 
in the next section).
One may then study the dynamics of a nonlinear system using the spectral
properties of this linear operator~\cite{koopmanPNAS1932}: for instance,
von~Neumann used this perspective in his celebrated proof of the mean ergodic
theorem~\cite{Neumann:1932a}.
Owing to its applicability in modeling nonlinear systems and to its demonstrated success in 
analyzing complex fluid flows, DMD has gained increasing popularity in fluid mechanics and beyond.
For instance, DMD has been utilized in the fields of epidemiology~\cite{proctorIH2015}, 
medicine~\cite{bourantasMP2014}, 
neuroscience~\cite{bruntonARXIV2014}, power systems~\cite{budisicCHAOS2012},  
robotics~\cite{bergerAR2015}, 
sustainable buildings~\cite{budisicCHAOS2012}, 
and video processing~\cite{grosekARXIV2014}.

Despite increasing adoption as a modeling and analysis tool, the adverse influence of 
measurement errors on DMD's performance and reliability remains under-appreciated.
For instance, the signal-to-noise ratio of the observed snapshot data can alter the 
growth/decay rates predicted by DMD~\cite{dukeEIF2012}---an obvious problem for studies that rely 
upon DMD to identify and distinguish between stable and unstable spatial modes.
Numerous other studies have also encountered DMD's sensitivity to measurement errors (e.g.,~sensor 
noise),  which has led to a host of approaches aimed at mitigating noise-related effects via 
various forms of rank-reduction, ensemble averaging, cross-validation, and 
windowing~\cite{dukeJFM2012,hematiPOF2014,schmidJFM2010,schmidEIF2011,schmidTCFD2011,semeraroEIF2012,tuJCD2014,wynnJFM2013}. 
As we will show, although these techniques provide a means of uniquely determining a DMD 
realization, the resulting analysis will be subject to systematic bias errors when the measured 
snapshot data are inexact due to sensor noise or other effects.
By viewing DMD as a ``best-fit'' least-squares/minimum-norm operator determined from measured 
snapshot data, we establish that the same sources of noise-induced bias arising in standard 
least-squares problems---extensively studied in statistics and numerical 
analysis~\cite{fierroNLAA1995,fierroLAIA1996,fierroSIAMJSC1997,golubJNA1980,golub1996,markovskySP2007,vanhuffel1991,zoltowskiSPIE1988}---will also plague DMD.

In this manuscript, we address the issue of noise-induced bias by focusing on the DMD algorithm directly.
We show that the currently used formulation of DMD accounts for errors in only
some of the snapshots, whereas measurement noise typically influences \emph{all} snapshots.
Invariably, accounting for noise in only a subset of the data will lead to biases, since doing so amounts to treating the remaining data as exact.
To arrive at an unbiased result, we propose a \emph{total-least-squares}/\emph{error-in-variables} formulation of DMD, such that errors in \emph{all} the data are considered.

In order to develop an unbiased noise-aware method, we rewrite DMD as a two-step procedure: (1) a subspace projection step and (2) an operator identification step.
In this form, it becomes simple to show that the subspace identification step
introduces a systematic error in existing DMD algorithms when the snapshot data are
inexact due to sensor noise and other factors.
We propose a slight modification to the conventional subspace identification step, based on an \emph{augmented snapshot matrix}, in order to remove the source of bias that is systematically introduced into current formulations of DMD.
The resulting noise-aware \emph{total DMD} (TDMD) framework reduces to the standard DMD algorithm when the measured snapshots are without error.
Moreover, based on the two-step analysis developed here, the de-biasing procedure is generalizable to other \emph{DMD-like} algorithms that appeal to Koopman spectral analysis, such as optimal mode decomposition~\cite{goulartCDC2012,wynnJFM2013}, streaming DMD~\cite{hematiPOF2014}, sparsity-promoting DMD~\cite{jovanovicPOF2014}, and optimized DMD~\cite{chenJNS2012}; in such cases, the ``operator identification'' step is to be replaced by the dynamical analysis algorithm of choice.

As we will show, even with imprecise snapshot measurements, TDMD successfully converges to the ``exact'' spectra for a simple linear system and for numerically simulated flow over a cylinder.
Furthermore, TDMD outperforms standard DMD in extracting dynamical information
from time-resolved particle image velocimetry~(TR-PIV) data of a separated flow.
We note that our focus here is on measurement noise and data quality, with particular
attention on removing the influence of such factors from the ensuing analysis.
While TDMD provides a systematic framework for conducting unbiased Koopman spectral analysis
in the context of \emph{measurement} noise, further investigation is needed to ensure that
such procedures do not remove system-specific \emph{process} noise, characterizations of which 
can provide descriptive physical insights.
An extensive discussion of Koopman spectral analysis for systems exhibiting weak random forcing
in the form of process noise can be found in~\cite{bagheriPOF2014}.
The influence of process noise on the Koopman spectrum, as characterized in~\cite{bagheriPOF2014},
together with noise-aware techniques like TDMD will be essential to disambiguating the 
contribution of various noise sources (i.e.,~data quality versus intrinsic stochastic forcing) on 
the resulting analysis.
%
%
Ultimately, TDMD provides a systematic framework for conducting unbiased Koopman spectral 
analysis in applied settings for which data quality can be an issue;
this will be essential for modeling complex systems and extracting credible dynamical descriptions 
from measured data.

\section{An Unbiased Formulation of DMD}
Consider a dynamical system given by $x\mapsto f(x)$, where $x\in\statesp$ is
the state variable.
This evolution law can be expressed in terms of the evolution of an appropriate set of
scalar-valued functions of state-space $g:\statesp\to\CC$, known as {\em observables}.
From this perspective, it is useful to consider the {\em Koopman operator}~$\koop$, an
infinite-dimensional linear operator that maps observables to corresponding
observables one step in the future:
$\koop g(x) = g(f(x))$~\cite{koopmanPNAS1931,mezicND2005}.
The utility here rests in the fact that the dynamics of the {\em nonlinear}\/ map~$f$ can be determined
completely from the {\em linear}\/ Koopman operator.

In recent years, analyzing practical systems via the spectral properties of the Koopman operator
(i.e., the eigenvalues, modes, and eigenfunctions of~$\koop$) has been made possible 
by means of \emph{DMD-like methods}~\cite{budisicCHAOS2012,mezicAnnRev2013,rowleyJFM2009,williamsAIMS2014,williamsPR2014}, 
in which one approximates the Koopman operator from data obtained from experiments or simulations,
without explicit knowledge of the map~$f$.
In these methods, one considers a vector of observables $u:\statesp\to\CC^n$ 
(typically $\Re^n$, in practice) evaluated at specific values $x_k\in\statesp$ 
and their images $f(x_k)$, for $k=1,\ldots,m$, and seeks a linear 
relationship between them:
\begin{equation}
  u(f(x_k))=Au(x_k),
  \label{eq:finitedimkoopman}
\end{equation}
where $A\in\Re^{n\times n}$. 
More specifically, the data consist of pairs of {\em snapshots}
$u(x_k), u(f(x_k))$, which may be obtained from an
experiment, for instance, by taking
measurements at two consecutive times.
Using the formalism in~\cite{tuJCD2014}, these snapshots are stored in the
$n\times m$ matrices
\begin{equation}
  \begin{gathered}
    X \defi \begin{bmatrix}u(x_1)&\cdots&u(x_m)\end{bmatrix}\\
    Y \defi \begin{bmatrix}u(f(x_1))&\cdots&u(f(x_m))\end{bmatrix},
  \end{gathered}
  \label{eq:snap}
\end{equation}
and from~\eqref{eq:finitedimkoopman}, one seeks a matrix~$A$ that satisfies
\begin{equation}
  Y = AX.
  \label{eq:dataeqn}
\end{equation}
In DMD, $A$ is given by the least-squares/minimum-norm solution to \eqref{eq:dataeqn}:
\begin{equation}
  \admd \defi {Y}{X}\pinv,
  \label{eq:dmddef}
\end{equation}
where $X\pinv$ denotes the Moore-Penrose pseudoinverse of~$X$.
It is shown in~\cite{tuJCD2014} and~\cite{williamsAIMS2014} that, under certain conditions on the data and
observables, the eigenvalues of~$A$ correspond to eigenvalues of the Koopman
operator~$\koop$, and Koopman eigenfunctions and modes may be found from $A$ as
well.
In other words, the methodology above provides a means for conducting Koopman
spectral analysis of dynamical systems directly from snapshot data.

The above discussion on connections between DMD and the Koopman operator focuses on the {\em underconstrained}\/ case with ``perfect'' snapshot data, in which
\eqref{eq:dataeqn} is satisfied exactly, and \eqref{eq:dmddef} gives the
minimum-norm solution.
Indeed, this case is common in many situations with exact snapshot data: for instance, it holds
whenever the columns of $X$ (the snapshots) are linearly independent.

Here, we are interested in applying Koopman spectral analysis in practical contexts 
with imperfect and noisy snapshot data; 
as such, we are primarily interested in the {\em overconstrained}\/ case, in which
we have more snapshots than observables ($m>n$).
In this case, \eqref{eq:dmddef} represents the least-squares solution
\begin{equation}
  \label{eq:dmd_ls}
  \min_{A,\Delta Y} \|\Delta Y\|_F,\quad\text{subject to}\quad Y + \Delta Y = AX,
\end{equation}
where $\|\cdot\|_F$ denotes the Frobenius norm.
(Note that, if the minimizing~$A$ is not unique, then \eqref{eq:dmddef} selects
the solution of minimum norm.)
Now, assume the data measurements are corrupted by some noise, which we do not
know.
One interpretation of~\eqref{eq:dmd_ls} is then to view $Y$ as the ``noisy'' snapshots and $-\Delta Y$ as the
``noise''; DMD then finds a linear relationship between the snapshots~$X$
and the ``noise-free'' snapshots $Y + \Delta Y$.

With this interpretation, it is apparent that the snapshots in~$X$ and~$Y$ are
treated asymmetrically: if we account for noise in the
measurements~$Y$, then it seems one should also account for noise in the
measurements~$X$ and solve the {\em total-least-squares} problem
\begin{equation}
  \label{eq:dmd_tls}
  \min_{A,\Delta X,\Delta Y}\left\|
    \begin{bmatrix}
      \Delta X\\
      \Delta Y
    \end{bmatrix}
  \right\|_F,\quad\text{subject to}\quad Y + \Delta Y = A(X + \Delta X).
\end{equation}
This is the central idea we propose here.
We shall see that treating $X$ and~$Y$ asymmetrically, as in~\eqref{eq:dmd_ls},
introduces a {\em bias} in the eigenvalues of~$A$, even in the context of noisy
snapshot data; 
in contrast, if we account for noise in both $X$ and~$Y$ as
in~\eqref{eq:dmd_tls}, then the bias is removed.
In fact, as shown in~\cite{gleserAS1981} and~\cite{vanhuffelAutomatica1989}, 
under certain assumptions on the data and provided that an exact linear relationship 
\eqref{eq:dataeqn} between snapshots exists in the noise-free case, then, 
in the presence of noise, the total-least-squares
solution converges to the exact solution as the number of snapshots tends to infinity, 
whereas the least-squares solution does not.

In order to solve the total-least-squares problem~\eqref{eq:dmd_tls}, we appeal
to a projection operator perspective~\cite{fierroNLAA1995,fierroLAIA1996,zoltowskiSPIE1988}.
Note that~\eqref{eq:dataeqn} may be written equivalently as
$X\trans A\trans = Y\trans$, where $\trans$ denotes Hermitian transpose.
Then, the least-squares solution may be obtained by projecting onto the range of
$X^*$. 
Writing this projection as $\proj_{X\trans}$ (and noting that orthogonal
projections are self-adjoint), we see the least-squares solution $A_\text{ls}$
satisfies
\begin{equation}
  \label{eq:1}
  Y\proj_{X\trans} = A_\text{ls}X\proj_{X\trans}.
\end{equation}
(Of course, $X\proj_{X\trans}=X$, but we leave the projection in~\eqref{eq:1} for
analogy with the total-least-squares case discussed below.)
It is clear that the DMD solution~\eqref{eq:dmddef} satisfies this relation,
noting that $X\pinv X = \proj_{X\trans}$.  Hence, when the usual DMD algorithm
is applied to overconstrained data, one may interpret it as first performing the
projections
\begin{equation}
  \label{eq:proj_ls}
  \bar Y = Y\proj_{X\trans}, \qquad
  \bar X = X\proj_{X\trans} = X,
\end{equation}
and then finding the minimum-norm solution of $\bar Y = A\bar X$.

An analogous approach can be used to solve the total-least-squares 
problem~\eqref{eq:dmd_tls}.
First, construct the {\em augmented snapshot matrix}\/
\begin{equation}
  \label{eq:2}
  Z\defi
  \begin{bmatrix}
    X\\
    Y
  \end{bmatrix},
\end{equation}
and let $Z_n$ denote the best rank-$n$ approximation of $Z$ (in the Frobenius
norm).
Then, the solution $A_\text{tls}$ of~\eqref{eq:dmd_tls} satisfies
\begin{equation}
  \label{eq:3}
  Y\proj_{Z\trans_n} = A_\text{tls} X\proj_{Z\trans_n},
\end{equation}
where $\proj_{Z\trans_n}$ denotes the projection onto the range of~$Z\trans_n$;
this may be found from the singular value decomposition of $Z$, 
as shown in Step~1 of the algorithm outlined below.
The solution of the total-least-squares problem may thus be obtained by first projecting the data
\begin{equation}
  \label{eq:proj_tls}
  \bar Y = Y\proj_{Z\trans_n},\qquad
  \bar X = X\proj_{Z\trans_n},
\end{equation}
and then finding the minimum-norm solution of $\bar Y = A\bar X$.

Both the least-squares and the total-least-squares solution approaches described above
amount to a single two-stage procedure consisting of 
(1)~a~{\em subspace projection} step, followed by (2)~an {\em operator identification} step,
distinguished from one another by the details of the subspace projection step.
That is, in the least-squares formulation~\eqref{eq:proj_ls}, 
only the $Y$ matrix is ``corrected'' to account for noise, and the correction (projection)
depends only on~$X$; in the total-least-squares
formulation~\eqref{eq:proj_tls}, both $X$ and~$Y$ are ``corrected,'' and the
projection depends on both $X$ and~$Y$.
We shall see in the next section that the former approach introduces a bias in
the eigenvalues of~$A$ when noise is present, while the latter approach does
not.

It is worth noting that, if the data matrices do satisfy $Y=AX$ exactly for some
$A$ (i.e., if the data are ``noise free''), then $\bar X = X$ and $\bar Y = Y$,
for both \eqref{eq:proj_ls} and~\eqref{eq:proj_tls}.
(To see this, note that if $Y=AX$, then $\range(Y\trans)\subset\range(X\trans)$,
and hence $\range(Z\trans)=\range(X\trans)$, which has dimension at
most~$n$.  Thus, $\bar Y\trans = \proj_{Z_n\trans} Y\trans= Y\trans$ and
$\bar X\trans = \proj_{Z_n\trans}X\trans=X\trans$.)
In other words, in the absence of noise, both methods are equivalent.

The least-squares and total-least-squares problems~\eqref{eq:dmd_ls}
and~\eqref{eq:dmd_tls} arise when the original problem~\eqref{eq:dataeqn} is
overconstrained, with $m>n$.
However, another common case is when we expect the dynamics to
evolve on a low-dimensional subspace, say of dimension~$r<n$.
In such instances, an underconstrained problem, with $n>m$, can be interpreted as 
an overconstrained problem when viewed on the low-dimensional subspace, 
provided that $r<m$.
The usual approach in this situation is to determine 
a suitable low-dimensional subspace from the data,
for instance using principal component analysis, to project the snapshots onto this
subspace, and then to proceed with~\eqref{eq:dmddef}, where $X$ and~$Y$ now contain
the projected snapshots~\cite{schmidJFM2010}.
In the context of the total-least-squares problem, this approach may prove unsatisfactory, 
since a subspace determined from $X$ or $Y$ alone may retain traces of the noise contamination.
Instead, we determine the ``best'' r-dimensional subspace from 
a truncated SVD of the augmented snapshot matrix $Z$, as described below.

We emphasize that in expressing total-least-squares DMD as a two-step process,
the subspace projection step~\eqref{eq:proj_tls} can be interpreted as a
``pre-processing'' step to be used for de-biasing \emph{any} DMD-like algorithm.
Thus, the method may be used with standard implementations of, for example,
optimal mode decomposition~\cite{goulartCDC2012,wynnJFM2013}, streaming
DMD~\cite{hematiPOF2014}, sparsity promoting DMD~\cite{jovanovicPOF2014}, or
optimized DMD~\cite{chenJNS2012}.
For instance, the de-biased algorithm for standard DMD~\cite{tuJCD2014} proceeds
as follows:
\begin{enumerate}
\item Compute the singular value decomposition $Z = U\Sigma V\trans$, and retain the
  first $n$ columns of $V$, denoting them $V_n$.  Note that $\proj_{Z_n\trans}=V_nV_n\trans$.
\item Project the snapshot matrices, calculating $\bar X = XV_nV_n\trans$ and
  $\bar Y = Y V_nV_n\trans$.
\item Calculate the reduced singular value decomposition $\bar X~=~\bar
  U\bar\Sigma\bar V\trans$.
\item Determine the DMD matrix $\tilde A_\text{dmd} = \bar U^* \bar Y\bar V\bar\Sigma^{-1}$, which is related to the full DMD operator $\admd=\bar U \tilde{A}_\text{dmd}\bar U^*$.
\item The DMD eigenvalues $\lambda_i$ are eigenvalues of $\tilde A_\text{dmd}$,
  and the corresponding DMD modes are $\varphi_i=\bar U w_i$, where $\tilde A_\text{dmd}w_i =
  \lambda_i w_i$.
\item If desired, calculate the associated frequency and growth
  rate for mode~$i$, as $f_i=~\angle\lambda_i/(2\pi\delta t)$ and
  $g_i=~\log |\lambda_i|/\delta t$, where $\delta t$ refers to the time-shift
  between snapshots stored in $X$ and $Y$.
\end{enumerate}
(Note that if the dynamics are expected to evolve on an $r$-dimensional subspace with $r<n$, replace $n$ with $r$ everywhere in the algorithm above.)

Lastly, note that while the total-least-squares formulation makes
DMD more ``robust'' to noise---in the sense that the framework explicitly accounts for 
inexact data and does not systematically introduce bias errors when 
applied to noisy data---the formulation can also make the solution procedure less stable;
total least-squares problems are known to exhibit less stability than least-squares
problems, though more robust solution approaches have been developed~\cite{fierroSIAMJSC1997,golubJNA1980,vanhuffelAutomatica1989,vanhuffel1991}.
While the term \emph{noise-robust} is often used to describe (regularized) total least-squares 
problems in the literature, in the remainder, we choose to use the term \emph{noise-aware} to 
emphasize the need for algorithmic techniques with greater computational robustness
than may be afforded by the de-biased DMD procedure outlined above.
Such issues are outside the scope of this study, but are the focus of ongoing work.

To distinguish the unbiased formulation from standard DMD in the remainder of the manuscript, we refer to this noise-aware framework as \emph{total DMD (TDMD)}, owing to its relationship with total least-squares.
In the following sections, we demonstrate the effectiveness of TDMD on a series of large-scale dynamical systems with noise-contaminated snapshot data.

\section{DMD on a linear system}
\begin{figure*}
  \begin{center}
    \centerline{
      \subfloat[$m=100$]{\includegraphics[width=0.32\textwidth]{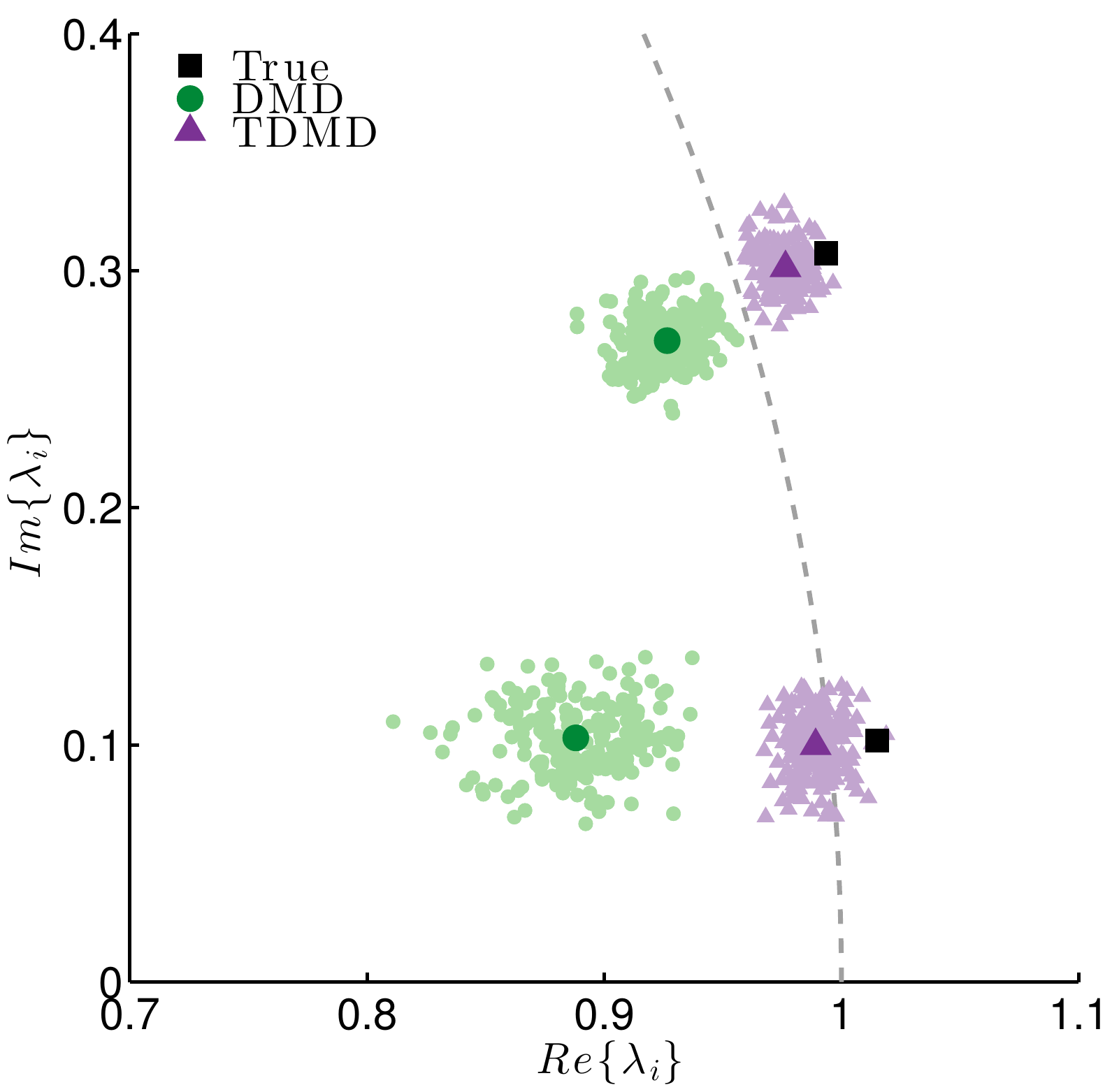}}\hfill
      \subfloat[$m=200$]{\includegraphics[width=0.32\textwidth]{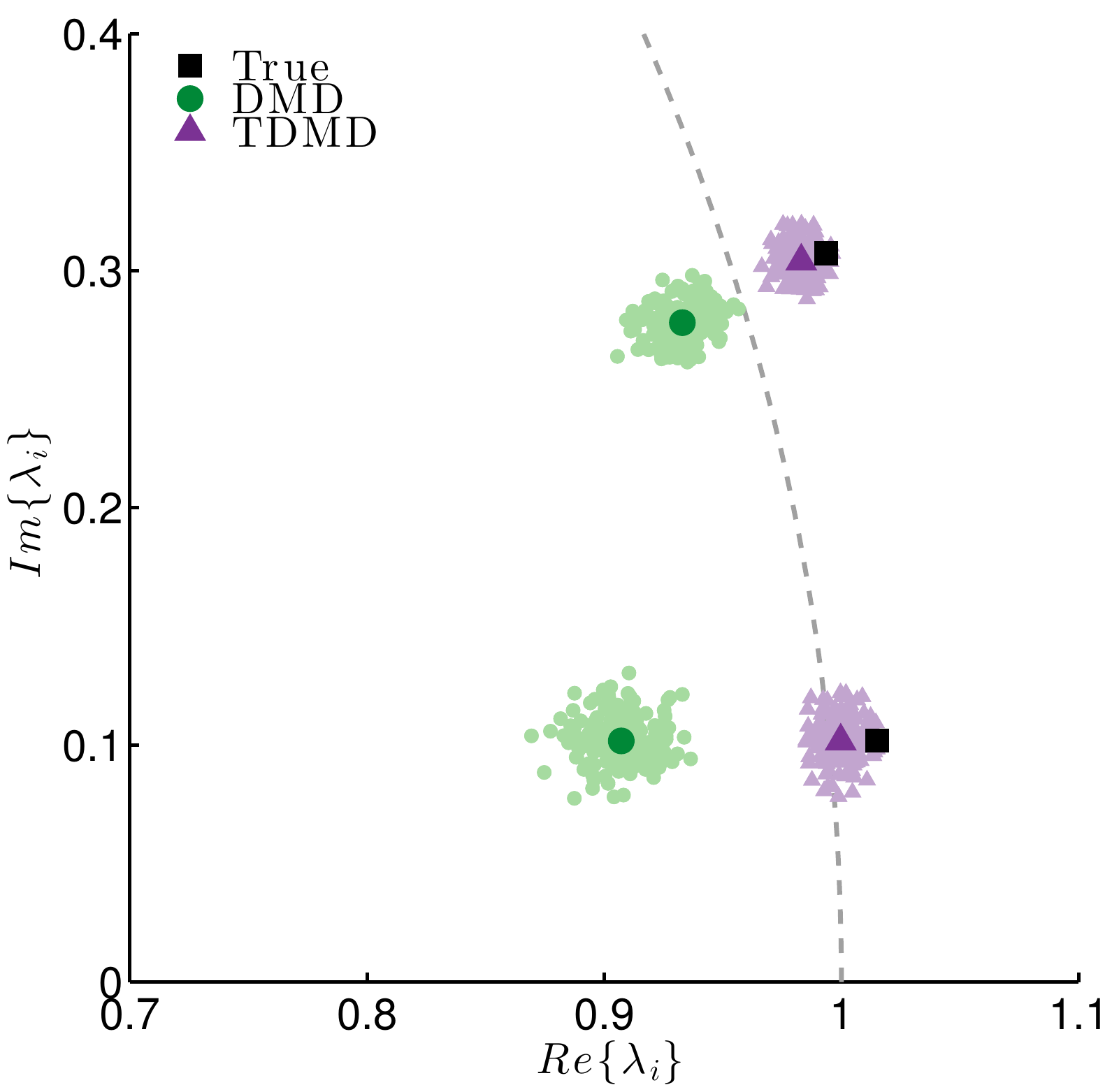}}\hfill
      \subfloat[$m=500$]{\includegraphics[width=0.32\textwidth]{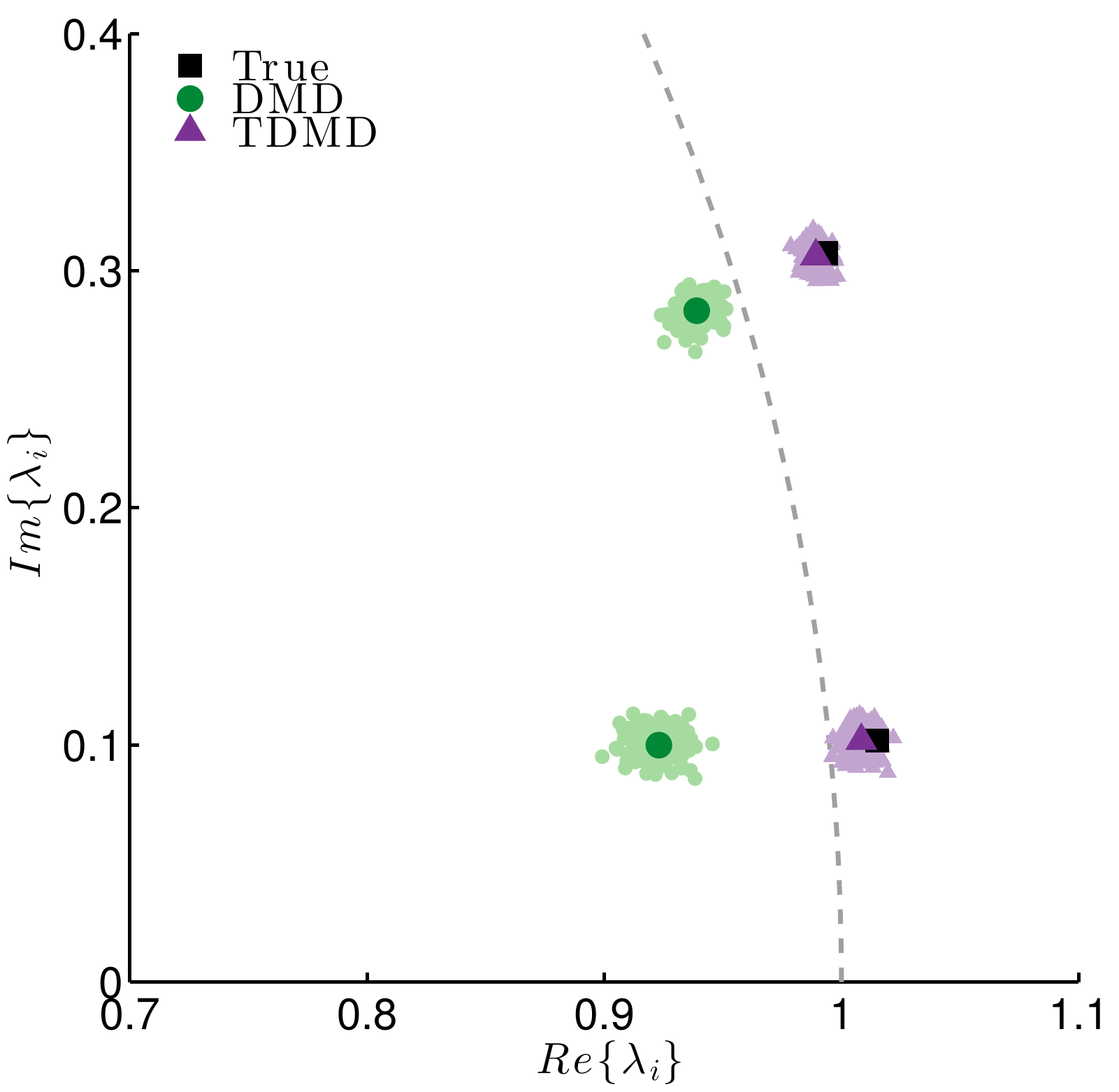}}\hfill
    }
  \end{center}
  \caption{\textbf{Spectrum for a Linear System.} The spectrum predicted by TDMD has essentially no bias and slightly tighter variance compared with standard DMD.  The simulated measurement noise is  $(\Delta X,\Delta Y)\sim\mathcal{CN}(0,0.05)$, and both methods set $r=2$.  The true eigenvalues (black squares) are plotted along with the mean values from standard DMD (dark circles) and TDMD (dark triangles) based on 200 different noise-realizations.  The eigenvalues from each of the individual realizations of standard DMD (light circles) and TDMD (light triangles) indicate the variance associated with each method.}
  \label{fig:measonly}
\end{figure*}

In order to highlight the ability of TDMD to yield unbiased approximations 
of the underlying dynamics in the context of noisy
data, we consider a ``toy problem'' for which the exact solution is known: 
a low-dimensional linear system with a large number of noisy observables.
In particular, we consider the state-space $\statesp~=~\CC^2$, with dynamics given
by $(x_1,x_2)\mapsto (\lambda_1 x_1, \lambda_2 x_2)$, with
$(\lambda_1,\lambda_2)~=~(1.02\textrm{e}^{0.1i},1.04\textrm{e}^{0.3i})$.  
The observable is a randomly chosen
linear transformation from $\CC^2$ to $\CC^{250}$ (i.e., $n=250$, $r=2$).
Snapshots are corrupted by additive circularly symmetric complex-valued Gaussian noise $(\Delta X,\Delta Y)\sim\mathcal{CN}(0,0.05)$.
Both standard and total DMD are performed on $m=\{100,200,500\}$ snapshot pairs.
In order to maintain a consistent signal-to-noise ratio between datasets in this comparison, 
we construct the snapshot data matrices by concatenating $\{5,10,25\}$ 
ensemble runs of 20 snapshots each, with each individual run initialized from a different state.
Each method is repeated for 200 independent 
noise-realizations of the data, 
and the resulting spectra are compared in Figure~\ref{fig:measonly}.
Even with a subspace projection to the known dimension of the underlying
dynamics ($r=2$), standard DMD yields a biased determination of the growth/decay characteristics;
in fact, for every realization, standard DMD \emph{erroneously} classifies both of the unstable modes as stable and decaying!
The frequencies identified by standard DMD possess a degree of bias as well.
This example highlights the potential pitfalls of previously employed ``noise-mitigation'' procedures such as ensemble averaging and cross-validation; DMD possesses bias in an expected value sense, so while such methods will reduce the variance, they will not remove the bias error.
In contrast, the unbiased TDMD formulation quickly converges to the correct spectrum, in an expected value sense, also with a decreasing variance as the number of collected snapshots increases;
thus, in this example, TDMD correctly classifies the modes as stable/unstable and predicts the associated frequencies correctly as well.
This suggests that commonly employed noise-mitigation techniques (e.g., ensemble averaging) can be applied with greater confidence in the TDMD setting.

\section{DMD on cylinder flow simulations}
\begin{figure*}
  \begin{center}
      \subfloat[$m=116$]{\includegraphics[width=0.32\textwidth]{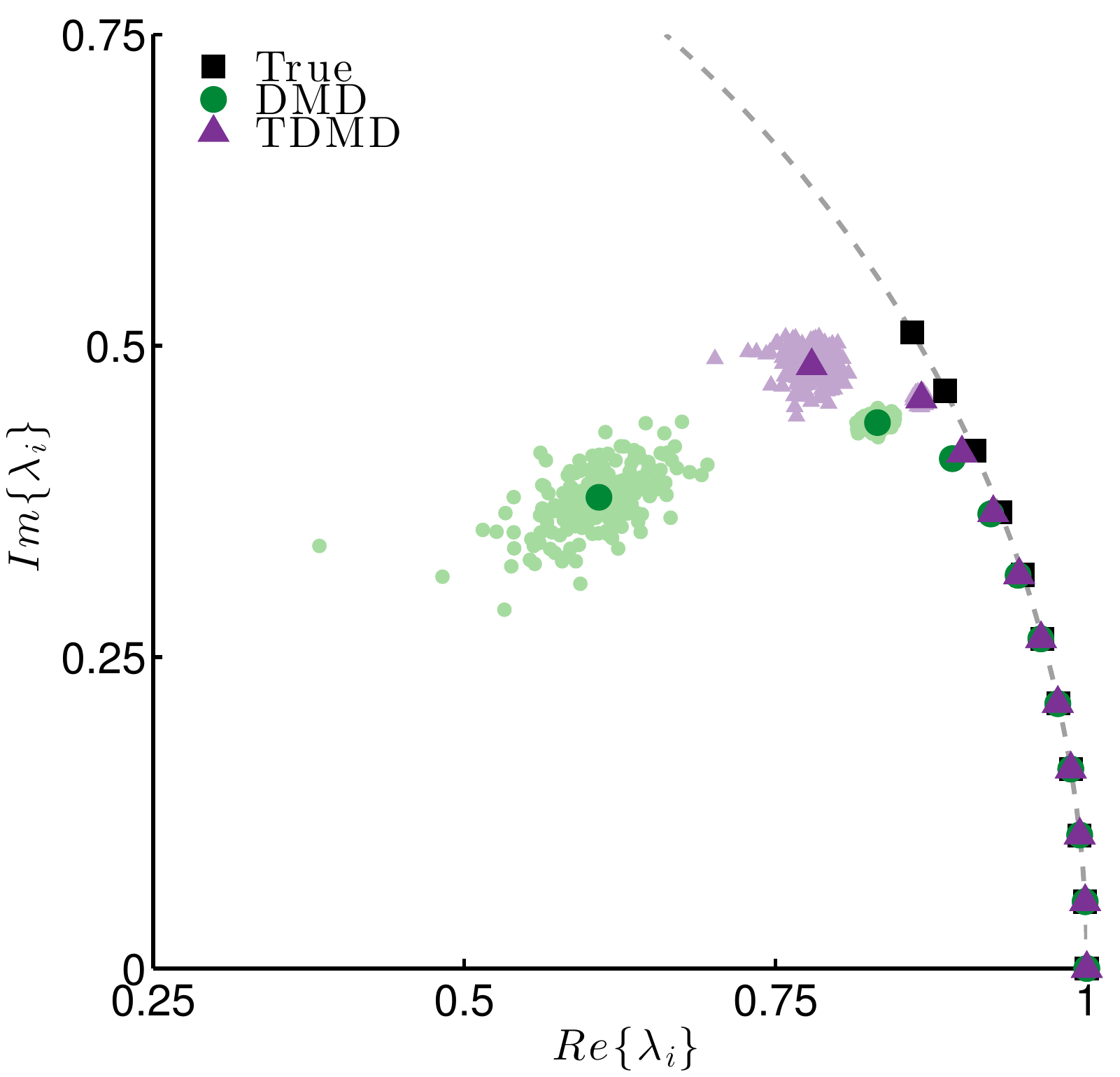}}
      \subfloat[$m=232$]{\includegraphics[width=0.32\textwidth]{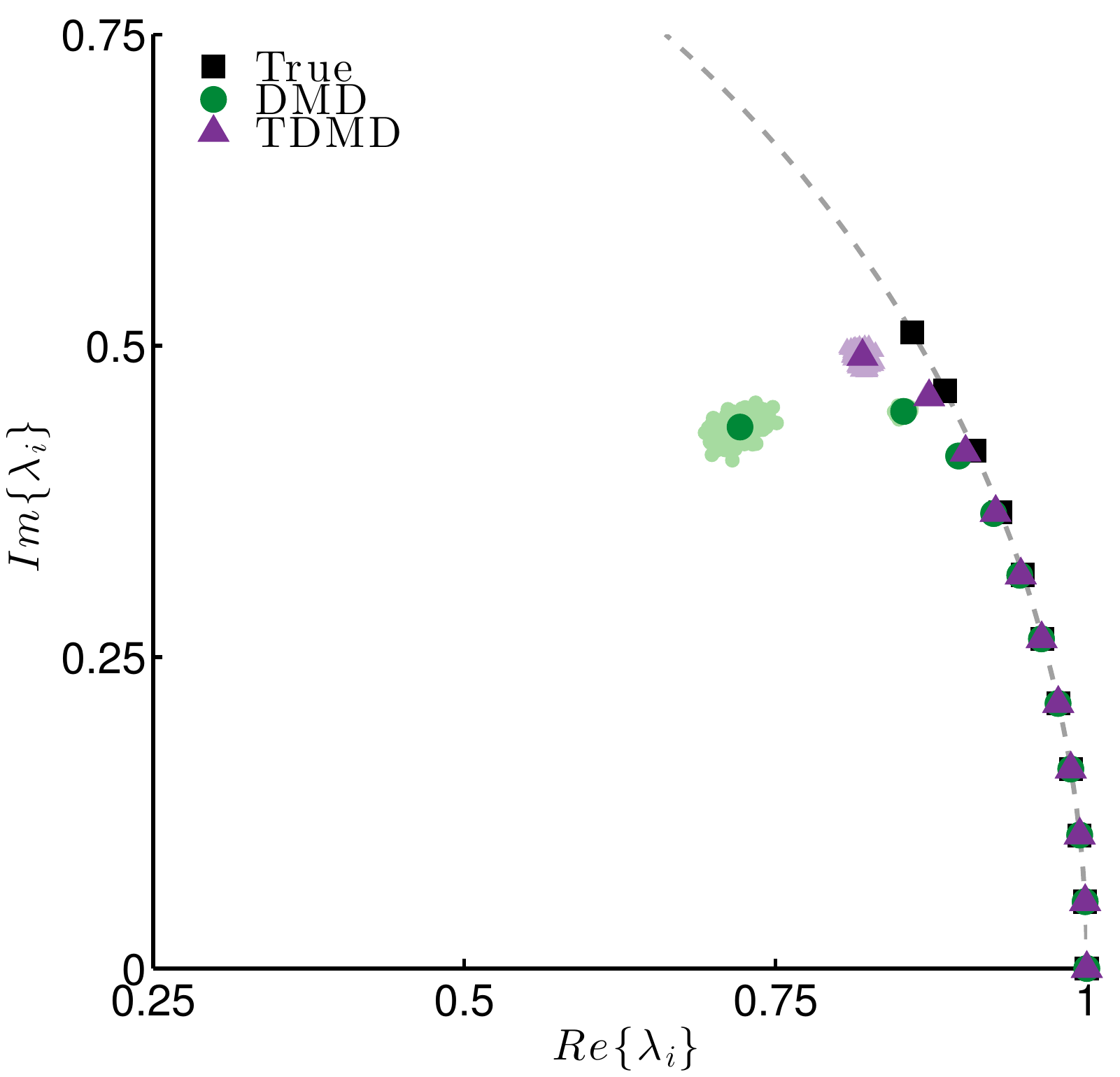}}
      \subfloat[$m=464$]{\includegraphics[width=0.32\textwidth]{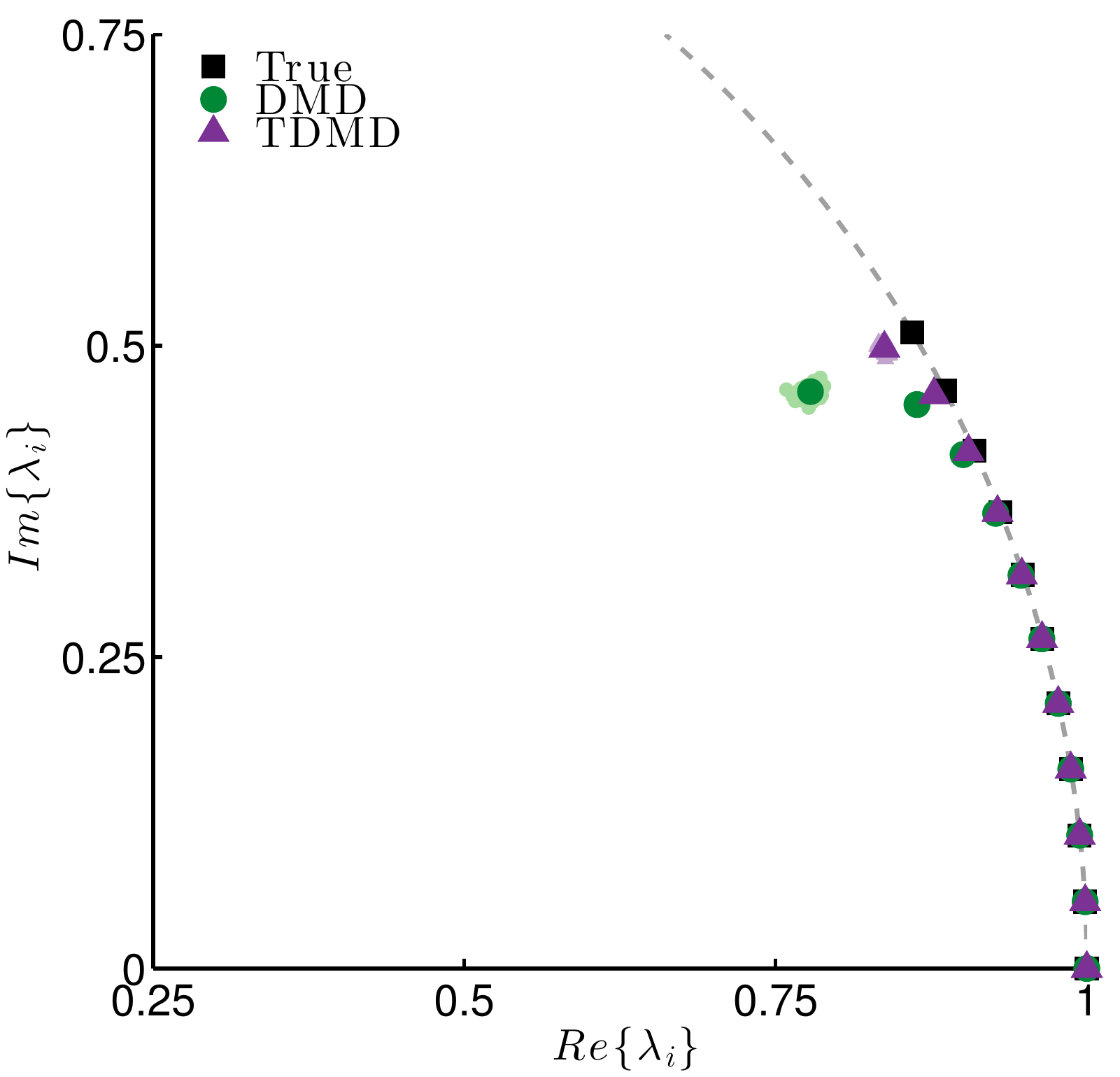}}
  \end{center}
  \caption{\textbf{Spectrum for the Flow over a Cylinder (DNS).} The spectrum
    predicted by TDMD converges to the true spectrum more quickly than standard
    DMD as the number of snapshots increases.  Here, the simulated measurement
    noise is $(\Delta X,\Delta Y)\sim\mathcal{N}(0,0.001)$, and both methods set
    $r=21$. The true eigenvalues (black squares) are plotted along with the mean
    values from standard DMD (dark circles) and TDMD (dark triangles) based on
    200 different noise-realizations.  The eigenvalues from each of the individual
    realizations of standard DMD (light circles) and TDMD (light triangles)
    indicate the variance associated with each method.}
  \label{fig:ibpm_mild}
\end{figure*}

While the demonstration of TDMD on a linear toy system showcases the advantages of the unbiased formulation over standard DMD in the simplest of cases, TDMD outperforms standard DMD in the analysis of more complex systems as well.
We establish the reliability of TDMD in the context of fluid flows by considering the canonical problem of flow past a cylinder.
Here, we aim to confirm the validity of TDMD in the context of mild noise contamination 
through the use of a simplistic model of noisy flowfield data; actual experimental 
datasets are considered in the next section.
Vorticity data, as reported in \cite{hematiPOF2014}, generated via direct numerical fluids simulations (DNS) and sampled at a rate of $f_s=100$Hz ($\delta t=0.01$s) are considered in this demonstration to ensure full control over (synthetic) measurement noise;
the Reynolds number based on cylinder diameter is $Re=100$.
To establish a baseline set of ``true'' DMD eigenvalues and modes, standard DMD is first applied to the set of ``exact'' snapshots (i.e., no noise corruption).
Next, the effect of measurement noise is considered by adding zero-mean Gaussian sensor noise $(\Delta X,\Delta Y)\sim\normal(0,0.001)$ to the exact vorticity snapshot data $(\bar{X},\bar{Y})$.
The effect of the number of snapshots $m$ is studied by concatenating the
original dataset ($n= 59\,501$, $m=116$) with itself, but with different realizations of additive measurement noise; here, cases with $m=\{116,232,464\}$ are considered.
The rank-reduction level $r=21$ is determined by seeking to retain over 99\% of the energy content based on the SVD of the noise corrupted data stored in $X$.
In this example, the computed DMD spectrum is not significantly altered by the noise---i.e., most of the DMD eigenvalues coincide with the ``true'' eigenvalues (see Figure \ref{fig:ibpm_mild}).
Even so, TDMD is able to handle the noise contamination more effectively; as the number of snapshots is
 increased, TDMD converges to the ``true'' spectrum more quickly than standard DMD.

\section{DMD on flow separation experiments}
TDMD's ability to extract the correct spectrum from synthetically corrupted numerical data garners trust for its use as a reliable method for fluid flow analysis; however, the assumptions of additive Gaussian measurement noise considered in our numerical study may be overly idealized.
A more compelling demonstration of TDMD's utility for noise-aware dynamical systems analysis 
can be made by working with noisy real-world data collected from a physical experiment.
As such, we now consider an experiment of separated flow over a flat plate 
(Re=$10^5$ with respect to chord length), with snapshots of the velocity field 
($n=42\,976$, $m=3\,000$) measured using  time-resolved particle image velocimetry~(TR-PIV) 
in a wind tunnel sampled at a rate of $f_s=1600$Hz ($\delta t=0.625$ms).
Further details of the experimental setup can be found in~\cite{griffinThesis2013}.
DMD and TDMD are performed using a rank-reduction level $r=25$, which corresponds to retaining over 99\% of energy content based on an SVD of $X$.
While the ``optimal'' truncation level is not necessarily the same for the two approaches,
we found $r=25$ allowed for a fair comparison on this dataset:
e.g.,~with $r=15$, TDMD does not yield any spurious eigenvalues and identifies some 
modes that are missed by DMD; 
alternatively, without truncation both methods give identical results.

\begin{figure*}
\begin{center}
\begin{minipage}[c][14cm]{0.7\textwidth}
\subfloat[Spectra]{\includegraphics[width=\textwidth]{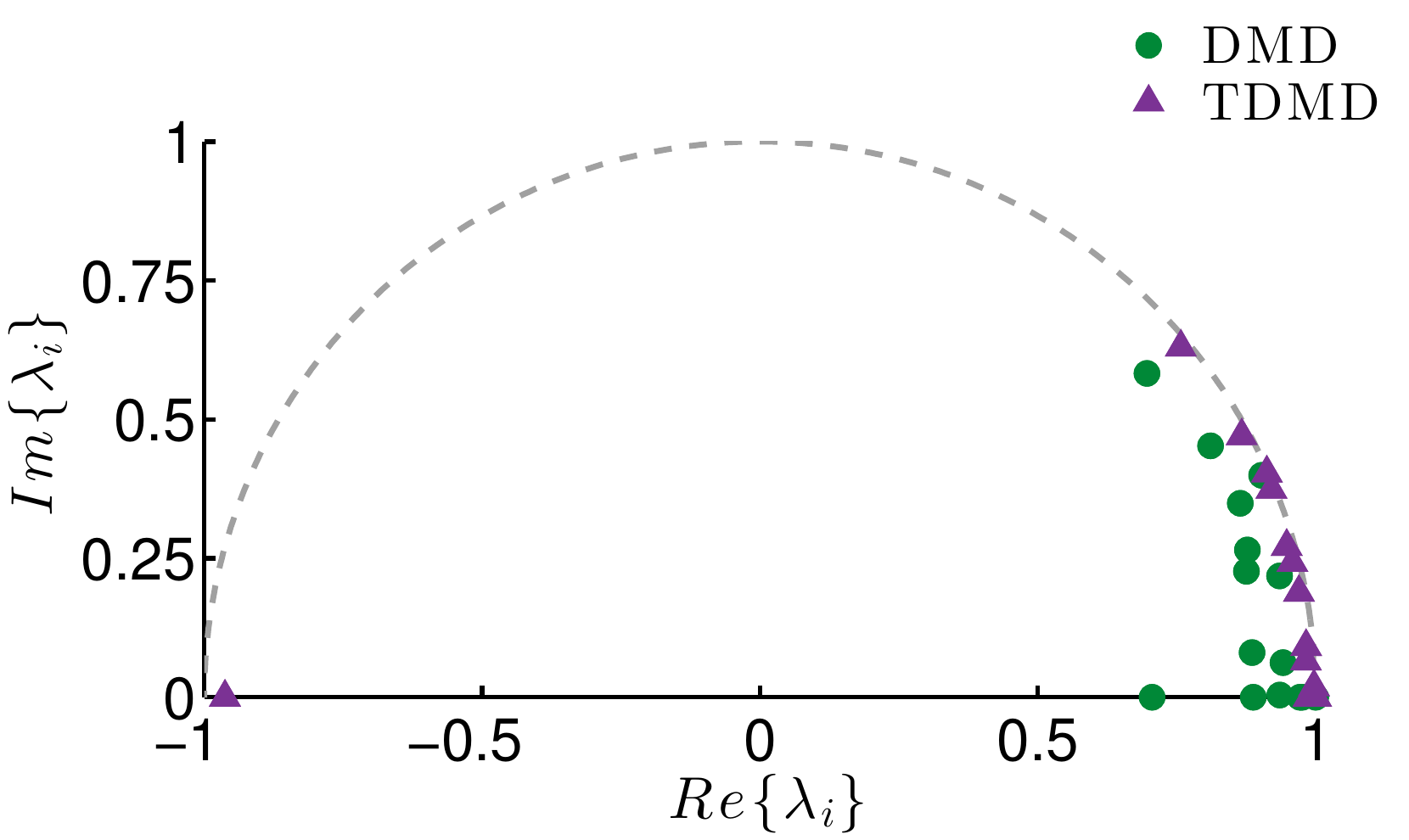}}\\
\subfloat[Mode Amplitudes]{\includegraphics[width=\textwidth]{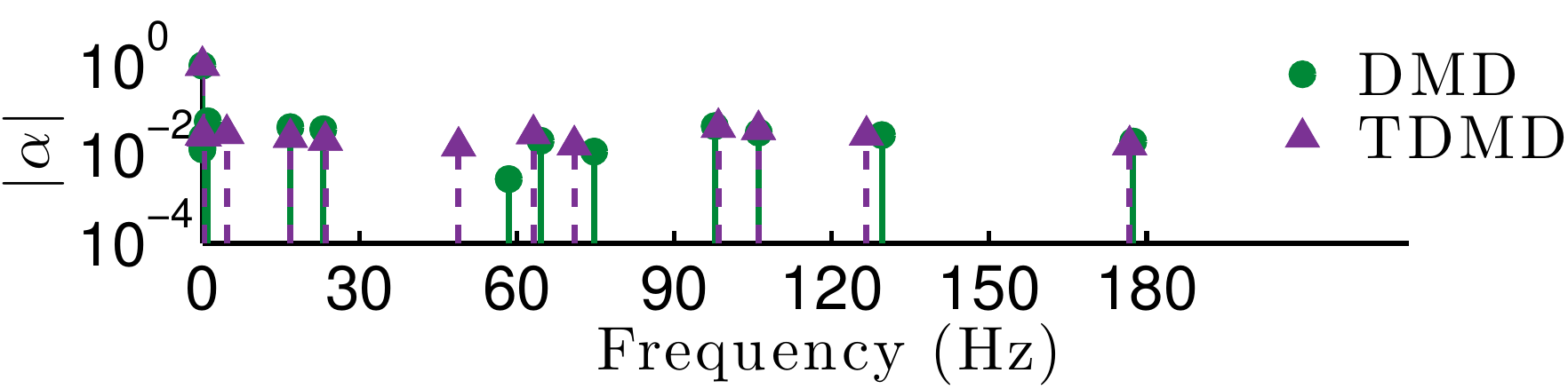}}
\end{minipage}
\caption{TDMD predicts smaller decay rates than DMD for the oscillatory modes that are extracted from TR-PIV data in a separated flow experiment.  
DMD and TDMD eigenvalues are plotted in (a) as circles and triangles, respectively.
Mode amplitudes are normalized by the maximum amplitude and plotted versus frequency in (b).}
\label{fig:pivspectra}
\end{center}
\end{figure*}

As seen in Figure~\ref{fig:pivspectra}(a), the dominant oscillatory modes extracted via TDMD 
are essentially non-decaying (i.e.,~eigenvalues have approximately unit magnitude), 
in contrast to those identified by standard DMD.
The mode amplitudes $|\alpha|$, plotted versus frequency in Figure~\ref{fig:pivspectra}(b), 
are computed with respect to the first snapshot in $X$ and normalized with respect 
to the maximum amplitude;
only ``dominant modes'' (i.e.,~$|\alpha|\ge 10^{-3}$) are reported.
A single spurious eigenvalue (with a negative real component) is reported by 
TDMD---an artifact of the ``de-regularizing'' nature of total-least-squares problems.
The amplitude of this spurious mode is very low compared to the other modes ($|\alpha|=10^{-5}$);
furthermore, for lower truncation levels, this spurious eigenvalue becomes more damped and eventually disappears,
but we keep $r=25$ for comparison with standard DMD.
Note that in practice, this spurious mode could be removed systematically by means of optimal 
mode selection techniques (e.g.,~sparsity promoting DMD~\cite{jovanovicPOF2014}).
Both standard DMD and TDMD identify similar frequencies, except in the range 30Hz--90Hz.
We note that TDMD identifies a mode at 49Hz, while standard DMD identifies a similarly shaped mode at 58Hz.
Local hot-wire measurements on the same flow configuration, previously reported 
in~\cite{griffinAIAA2013,griffinThesis2013}, corroborate the presence of a frequency 
peak in the range 45Hz--50Hz, as identified by TDMD.
%
%
In Figure~\ref{fig:pivmodes}, the real components of the dominant oscillatory vorticity modes---computed as the curl of DMD/TDMD velocity modes---are plotted.
Qualitative differences in the mode shapes corresponding to frequencies below 100Hz reveals a disparity in the dynamical characterization of the two methods (e.g., compare tiles (f) and (o) in 
Figure~\ref{fig:pivmodes});
these qualitative shape differences even arise for the modes with matching frequencies (e.g., compare tile (d) with (m), (h) with (q), and (i) with (r) in Figure~\ref{fig:pivmodes}).
%
%
%
Both methods display sensitivity to the truncation level $r$, indicating further investigation may be warranted;
still, qualitative differences arise between the two analysis methods for other truncation levels.
While TDMD demonstrates modest gains over standard DMD in this context, the results do demonstrate 
TDMD's ability to extract dynamical content that is consistent with other techniques.
On these grounds, TDMD seems to outperform standard DMD in extracting a more representative dynamical description of the system.

\newcommand{\tilesize}{0.25\textwidth}
\newcommand{\tilespace}{\hspace{100pt}}
\begin{figure*}
\begin{minipage}[c][21.5cm]{\textwidth}
\begin{center}
\subfloat{\textbf{DMD} \hspace{30pt}}\tilespace\subfloat{\ \hspace{70pt} \textbf{TDMD}}\\
    \setcounter{subfigure}{0}
\subfloat[$f~=~177~\text{Hz}$,~$|\lambda|~=~0.91$,~$|\alpha|~=~0.02$]{\includegraphics[width=\tilesize]{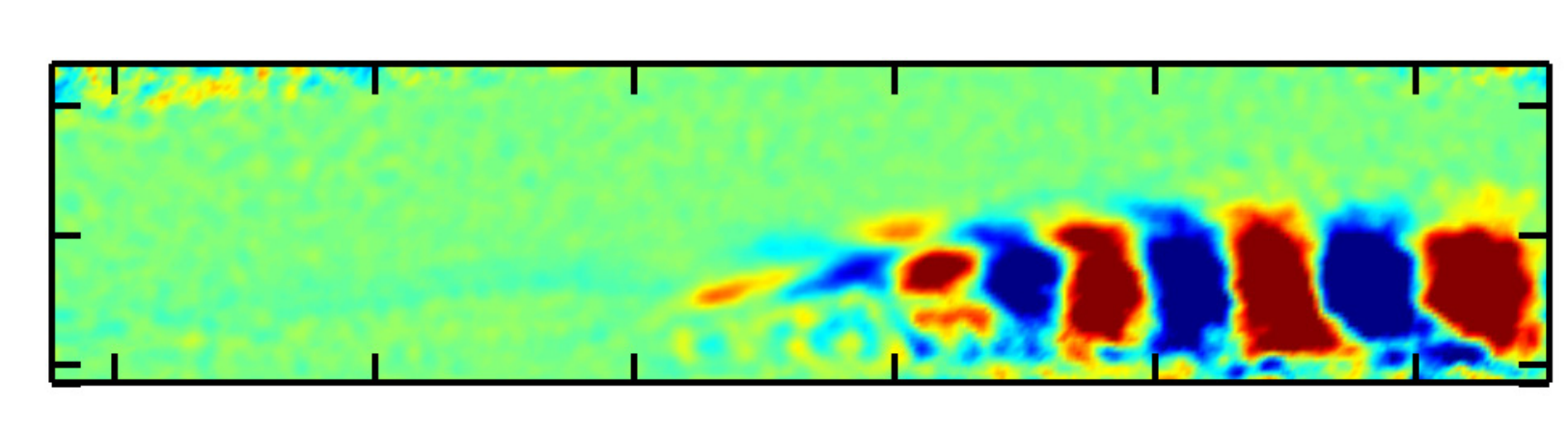}}\tilespace
    \setcounter{subfigure}{9}
\subfloat[$f~=~177~\text{Hz}$,~$|\lambda|~=~0.99$,~$|\alpha|~=~0.02$]{\includegraphics[width=\tilesize]{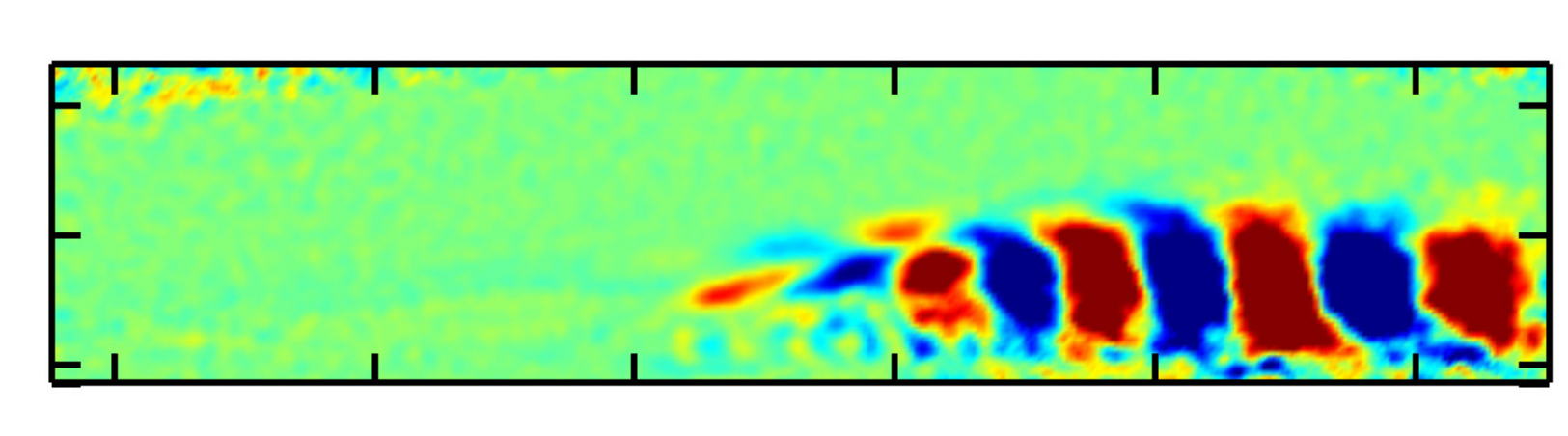}}\\
    \setcounter{subfigure}{1}
\subfloat[$f~=~130~\text{Hz}$,~$|\lambda|~=~0.93$,~$|\alpha|~=~0.03$]{\includegraphics[width=\tilesize]{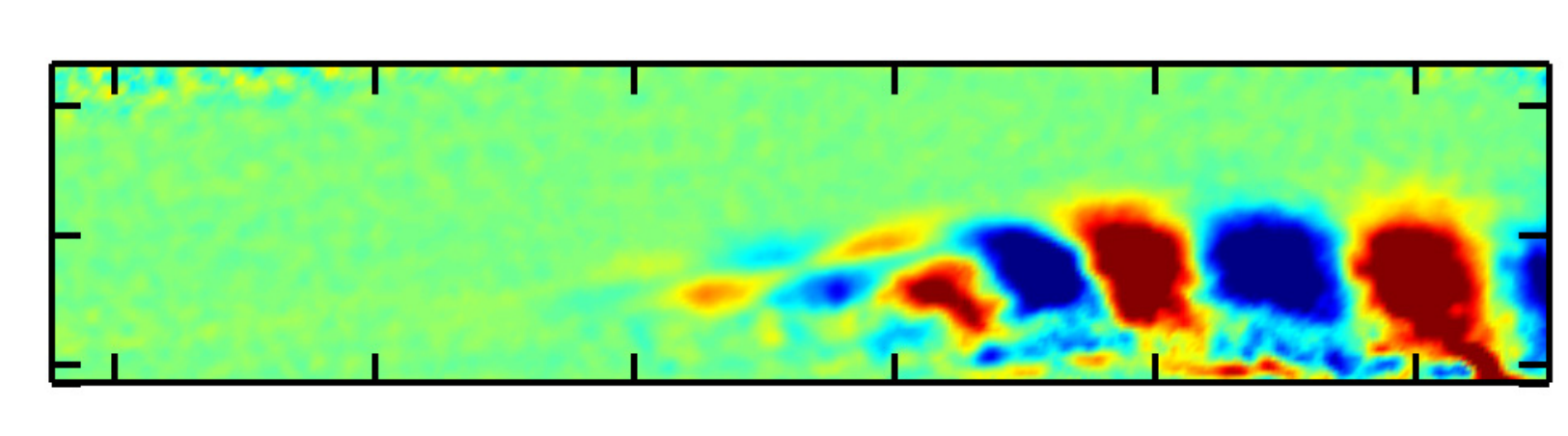}}\tilespace
    \setcounter{subfigure}{10}
\subfloat[$f~=~127~\text{Hz}$,~$|\lambda|~=~0.99$,~$|\alpha|~=~0.03$]{\includegraphics[width=\tilesize]{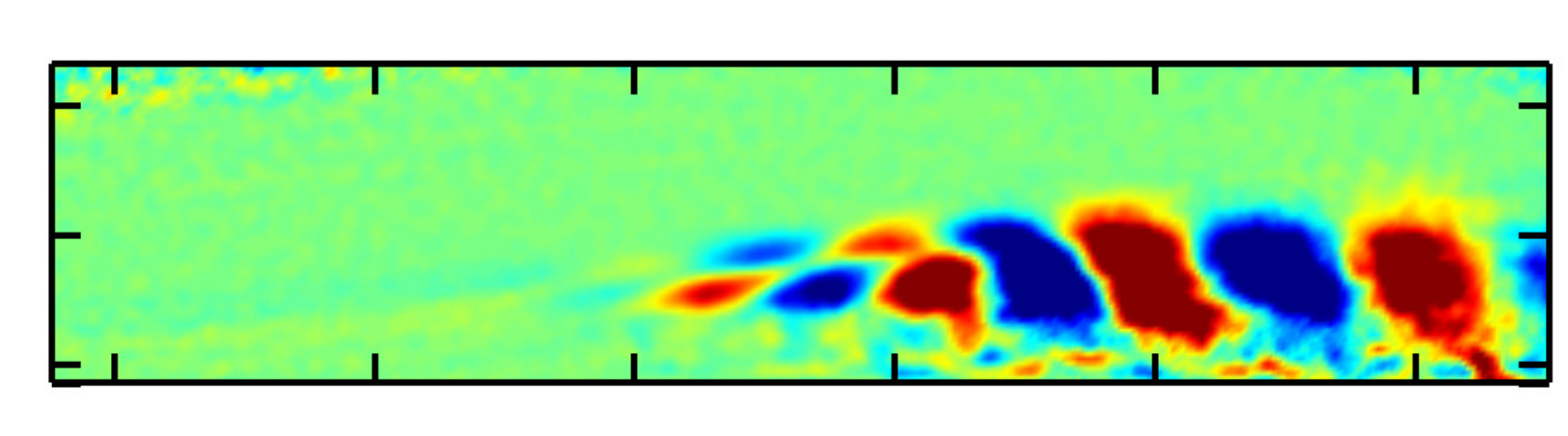}}\\
    \setcounter{subfigure}{2}
\subfloat[$f~=~106~\text{Hz}$,~$|\lambda|~=~0.99$,~$|\alpha|~=~0.03$]{\includegraphics[width=\tilesize]{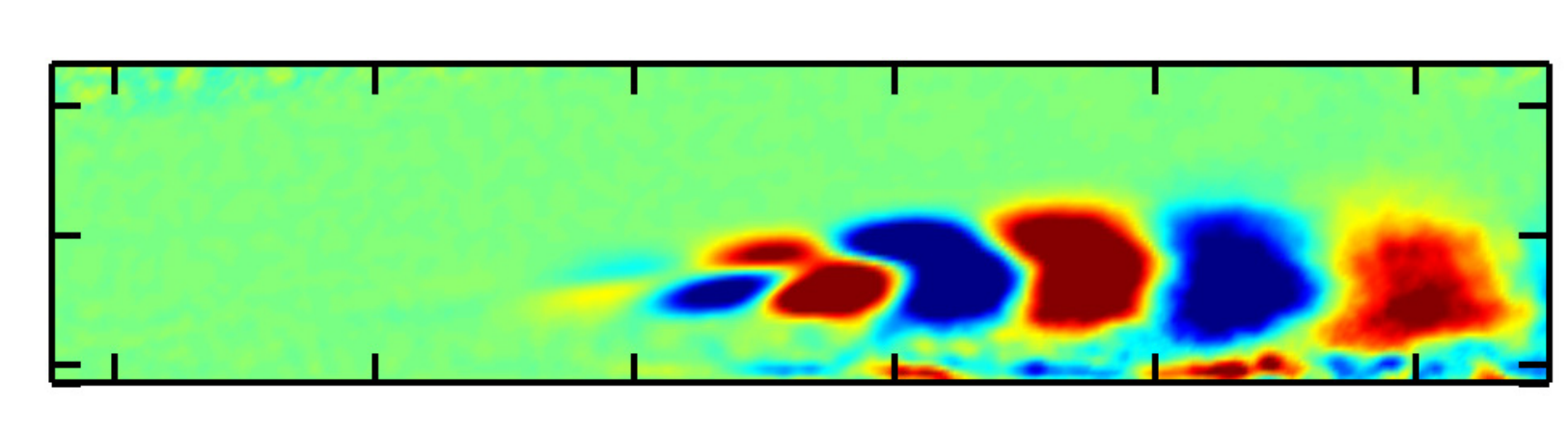}}\tilespace
    \setcounter{subfigure}{11}
\subfloat[$f~=~106~\text{Hz}$,~$|\lambda|~=~1.0$,~$|\alpha|~=~0.04$]{\includegraphics[width=\tilesize]{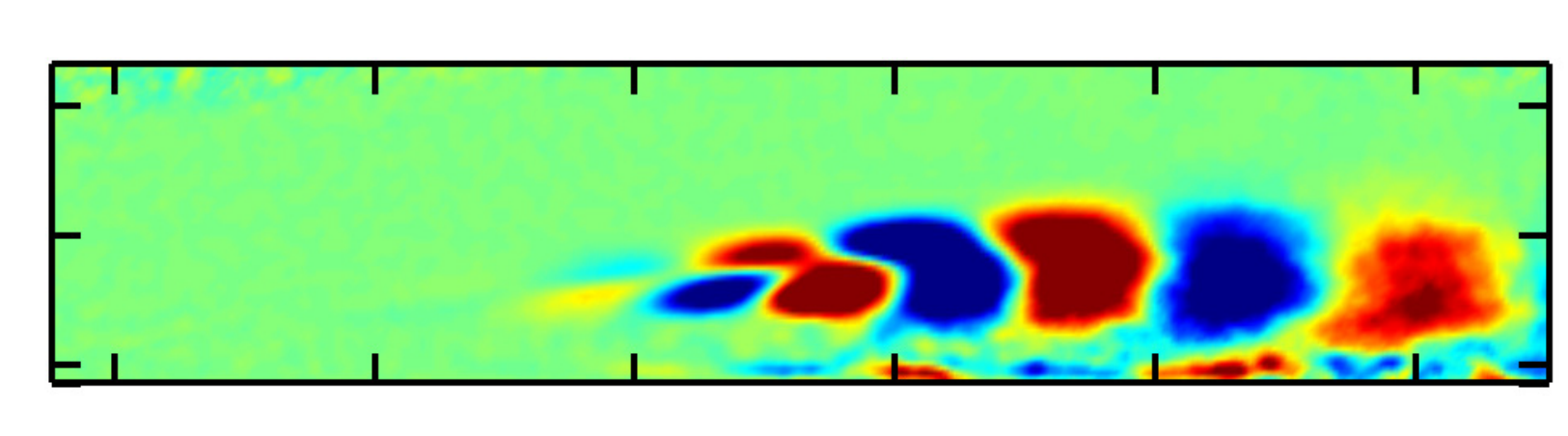}}\\
    \setcounter{subfigure}{3}
\subfloat[$f~=~98~\text{Hz}$,~$|\lambda|~=~0.93$,~$|\alpha|~=~0.04$]{\includegraphics[width=\tilesize]{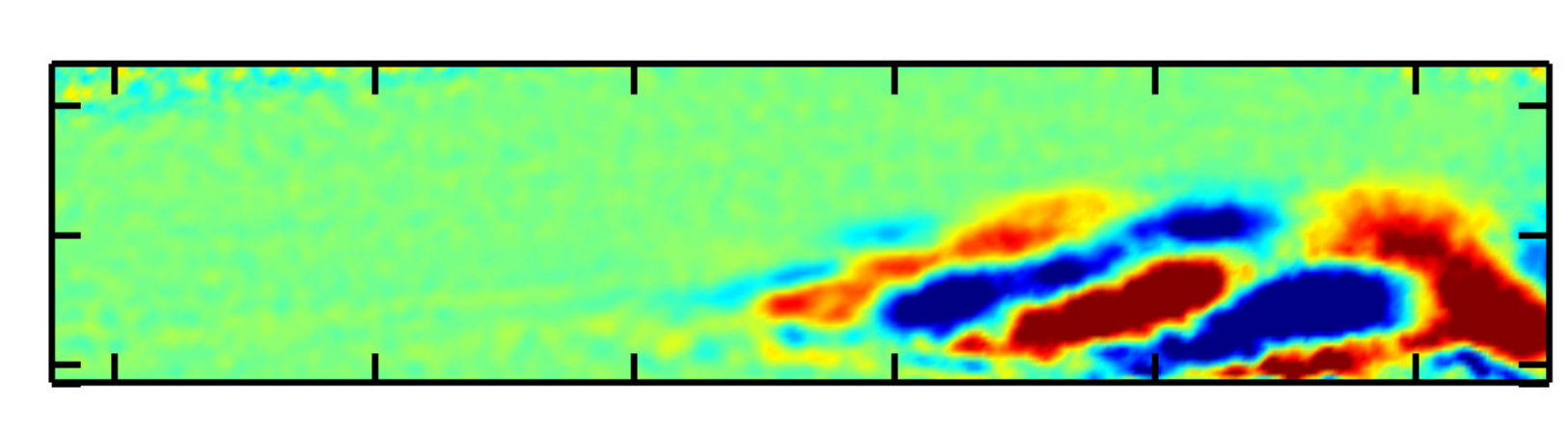}}\tilespace
    \setcounter{subfigure}{12}
\subfloat[$f~=~98~\text{Hz}$,~$|\lambda|~=~0.99$,~$|\alpha|~=~0.04$]{\includegraphics[width=\tilesize]{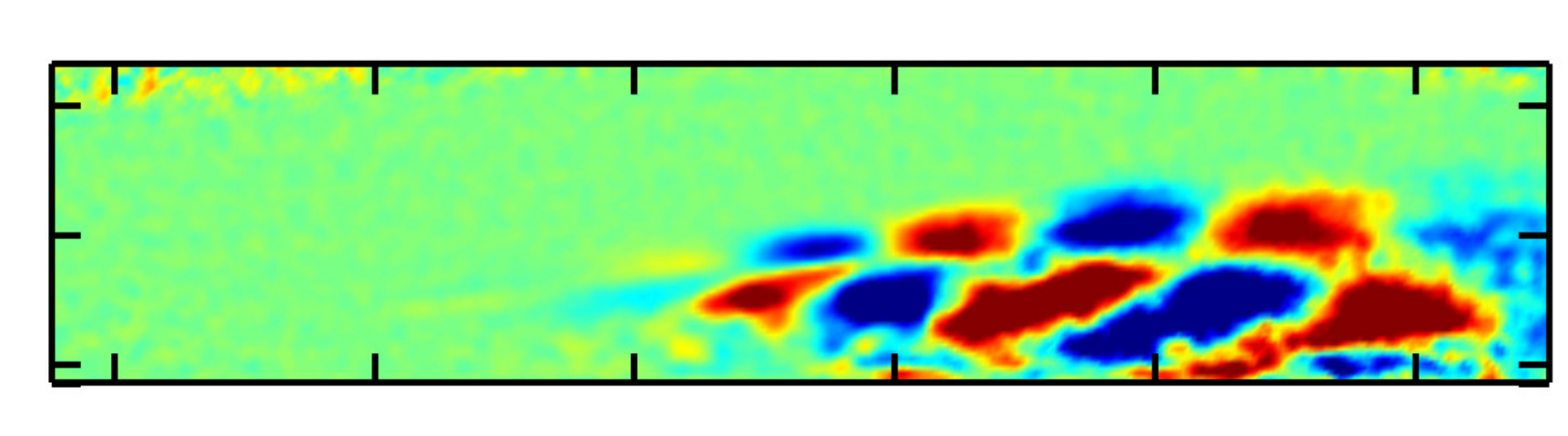}}\\
    \setcounter{subfigure}{4}
\subfloat[$f~=~75~\text{Hz}$,~$|\lambda|~=~0.92$,~$|\alpha|~=~0.01$]{\includegraphics[width=\tilesize]{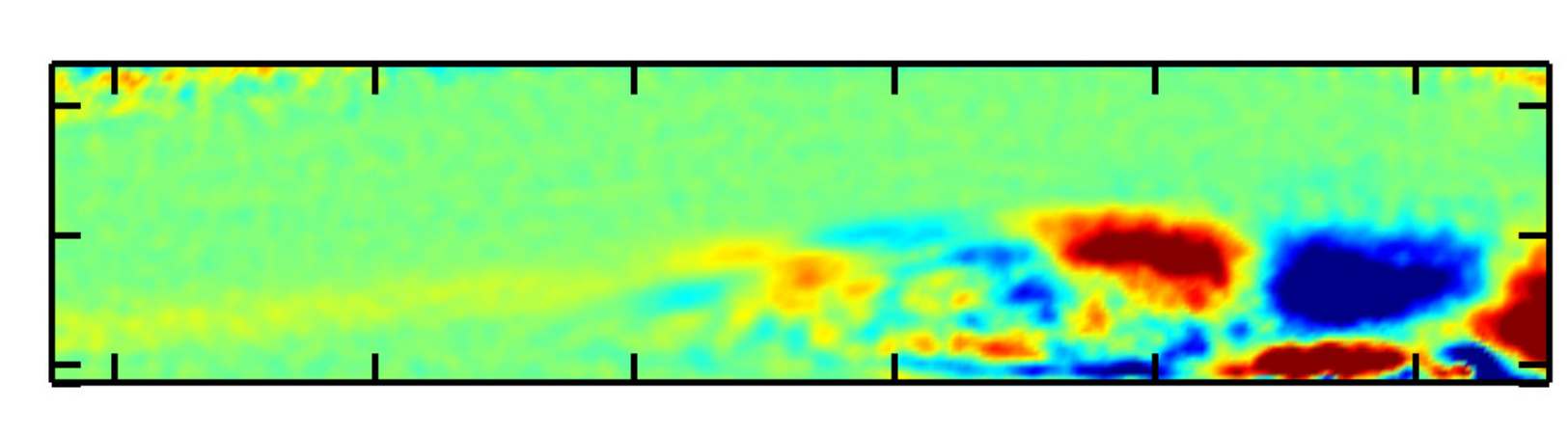}}\tilespace
    \setcounter{subfigure}{13}
\subfloat[$f~=~71~\text{Hz}$,~$|\lambda|~=~0.99$,~$|\alpha|~=~0.02$]{\includegraphics[width=\tilesize]{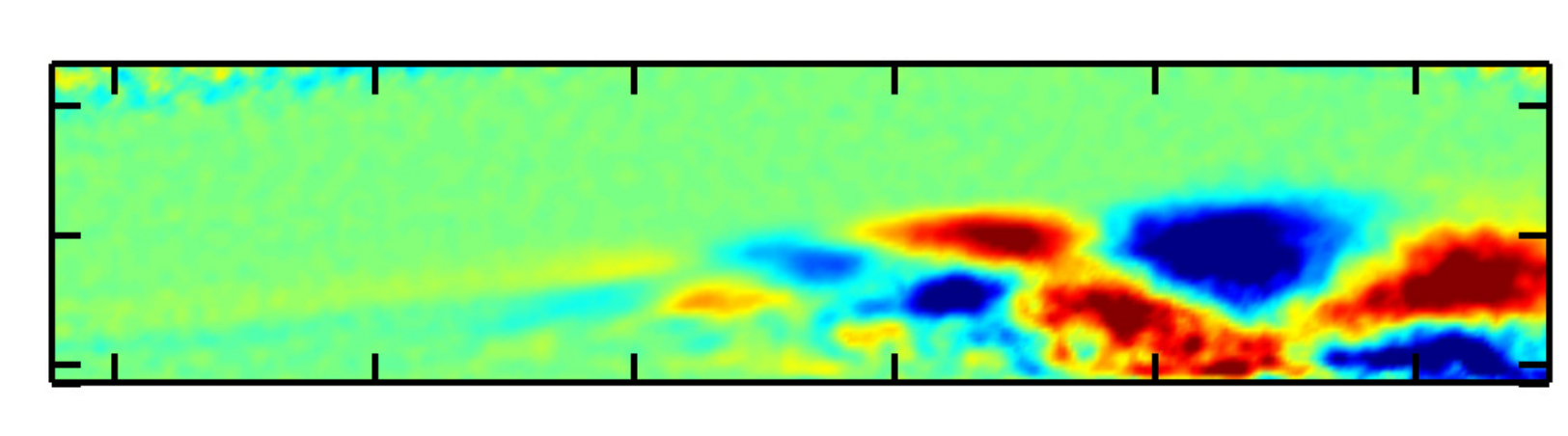}}\\
    \setcounter{subfigure}{5}
\subfloat[$f~=~65~\text{Hz}$,~$|\lambda|~=~0.90$,~$|\alpha|~=~0.02$]{\includegraphics[width=\tilesize]{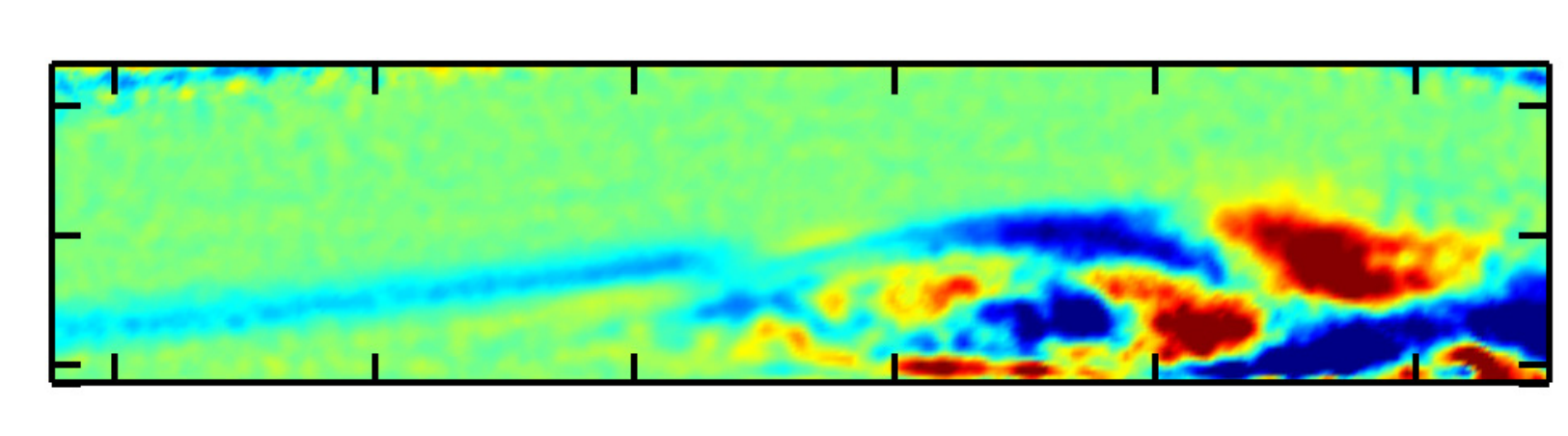}}\tilespace
    \setcounter{subfigure}{14}
\subfloat[$f~=~63~\text{Hz}$,~$|\lambda|~=~0.99$,~$|\alpha|~=~0.03$]{\includegraphics[width=\tilesize]{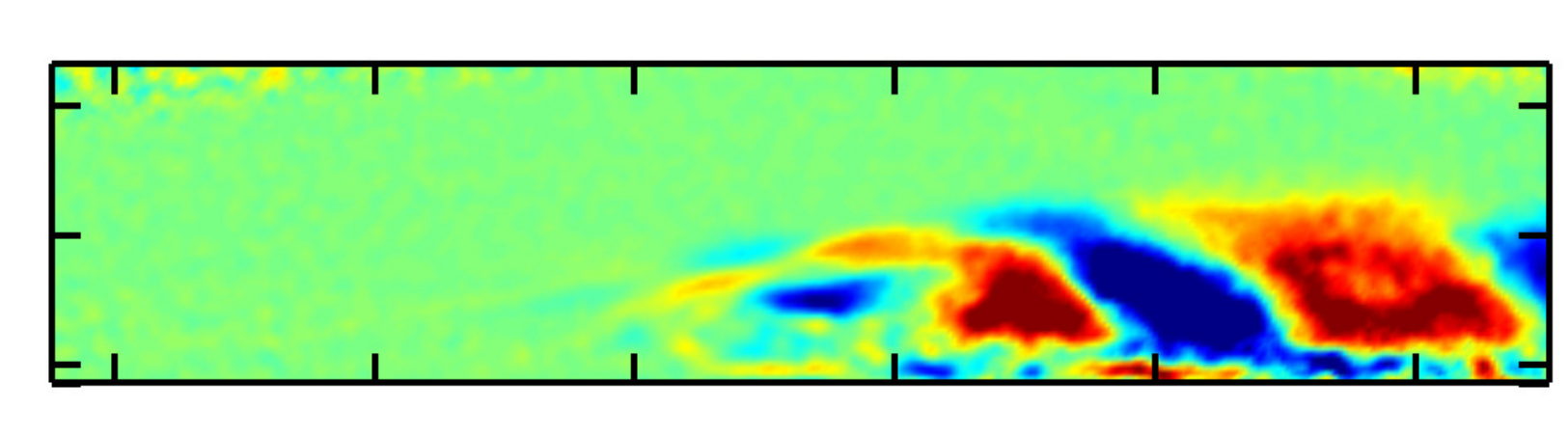}}\\
    \setcounter{subfigure}{6}
\subfloat[$f~=~58~\text{Hz}$,~$|\lambda|~=~0.96$,~$|\alpha|~=~0.003$]{\includegraphics[width=\tilesize]{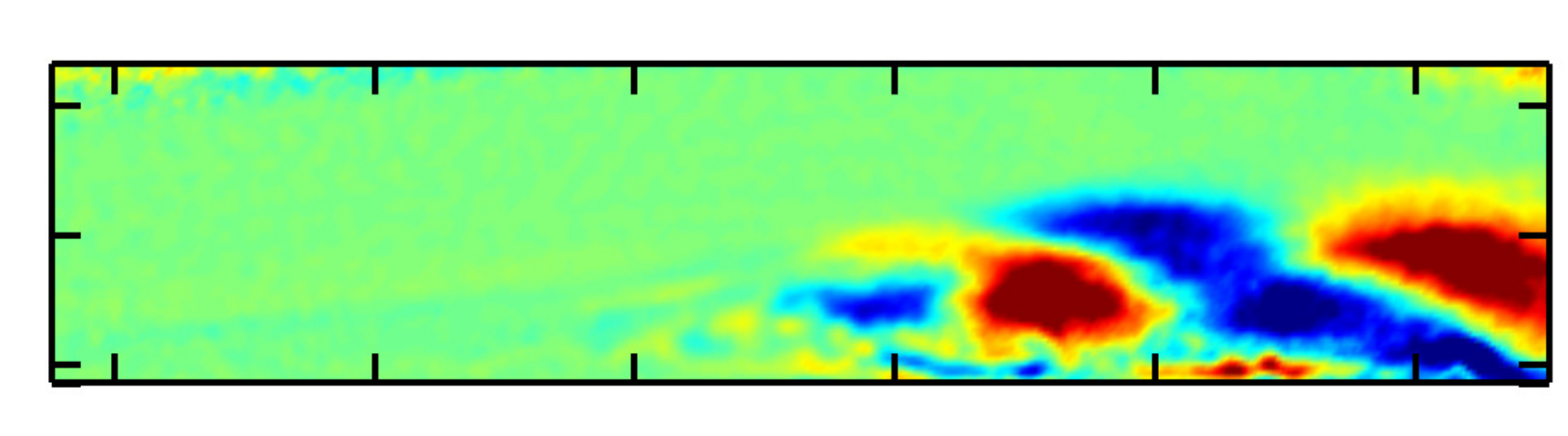}}\tilespace
    \setcounter{subfigure}{15}
\subfloat[$f~=~49~\text{Hz}$,~$|\lambda|~=~0.99$,~$|\alpha|~=~0.02$]{\includegraphics[width=\tilesize]{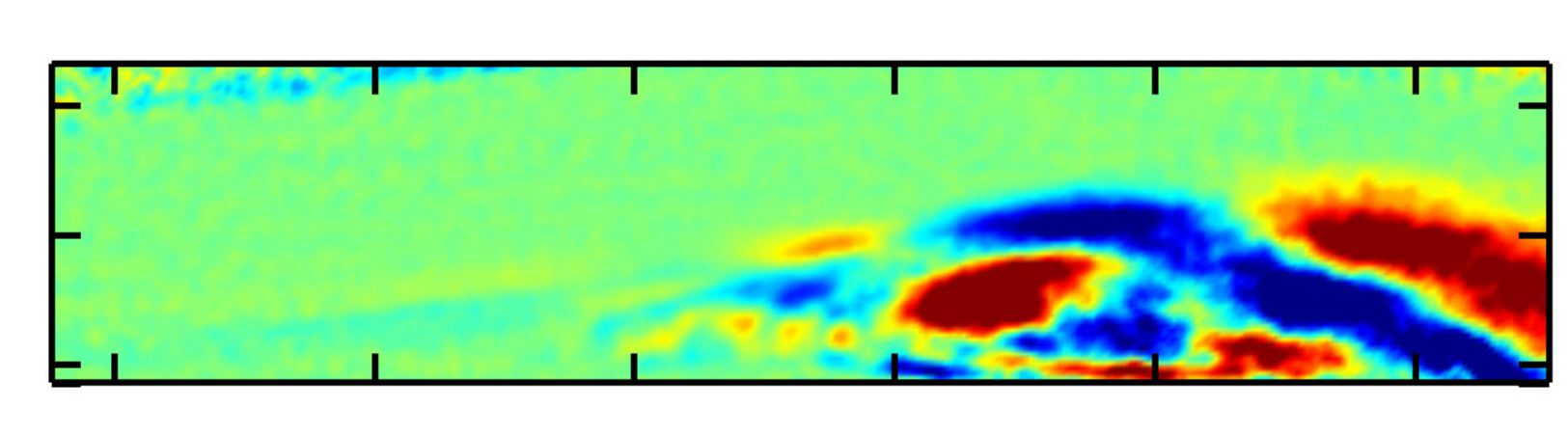}}\\
    \setcounter{subfigure}{7}
\subfloat[$f~=~23~\text{Hz}$,~$|\lambda|~=~0.89$,~$|\alpha|~=~0.04$]{\includegraphics[width=\tilesize]{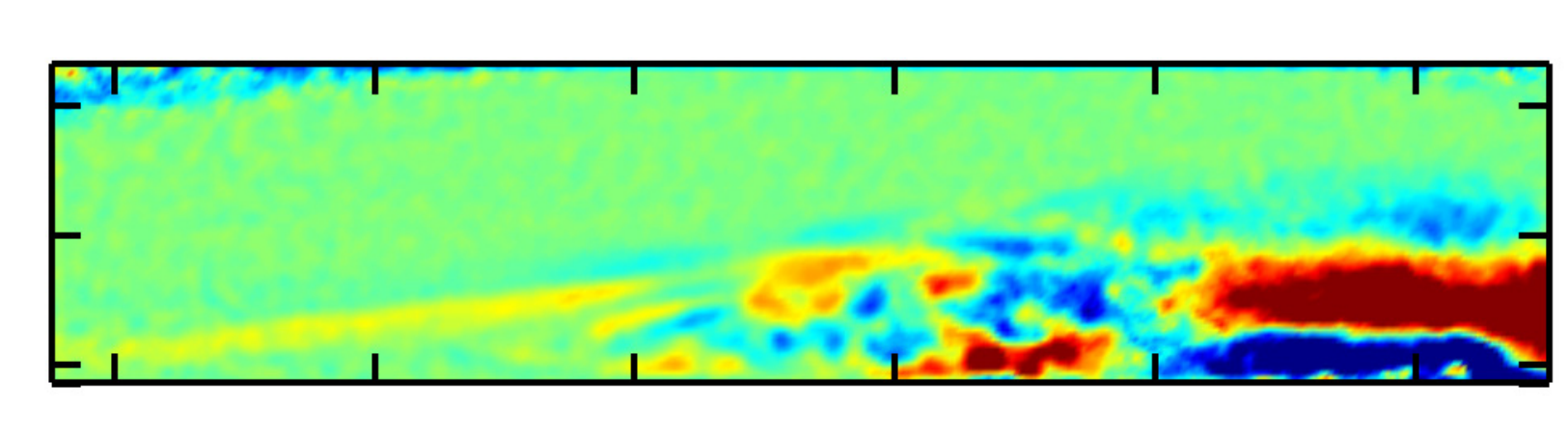}}\tilespace
    \setcounter{subfigure}{16}
\subfloat[$f~=~23~\text{Hz}$,~$|\lambda|~=~0.99$,~$|\alpha|~=~0.02$]{\includegraphics[width=\tilesize]{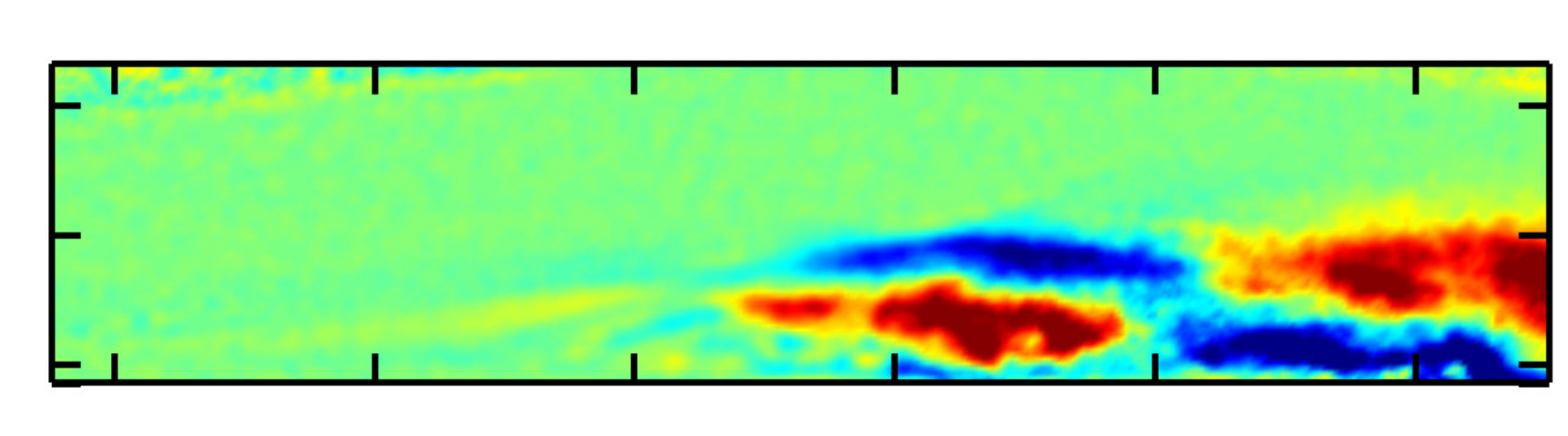}}\\
    \setcounter{subfigure}{8}
\subfloat[$f~=~17~\text{Hz}$,~$|\lambda|~=~0.94$,~$|\alpha|~=~0.04$]{\includegraphics[width=\tilesize]{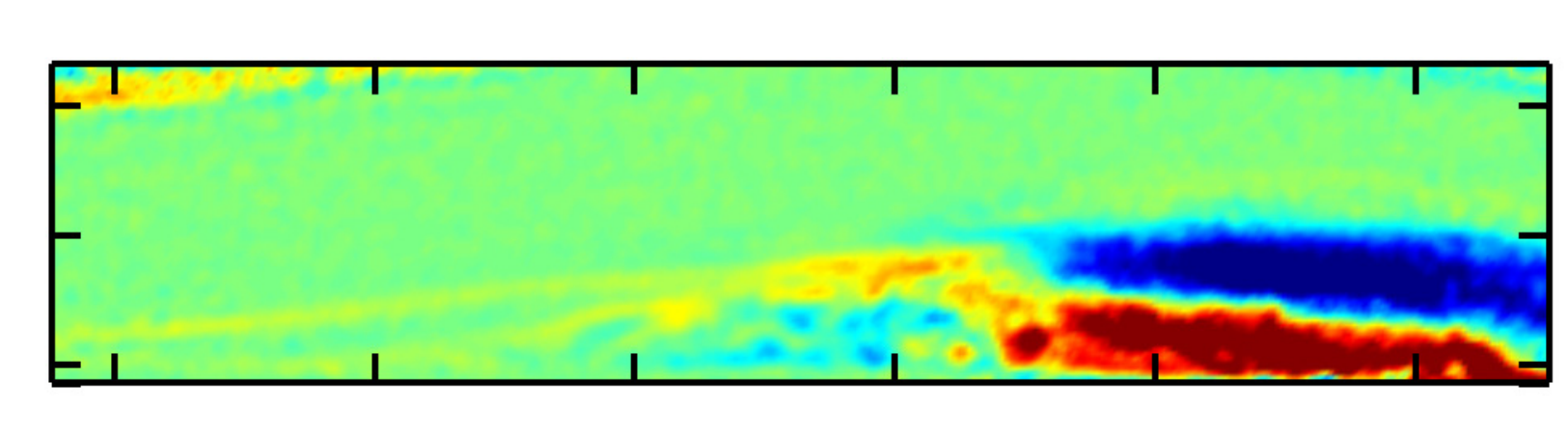}}\tilespace
    \setcounter{subfigure}{17}
\subfloat[$f~=~17~\text{Hz}$,~$|\lambda|~=~0.98$,~$|\alpha|~=~0.02$]{\includegraphics[width=\tilesize]{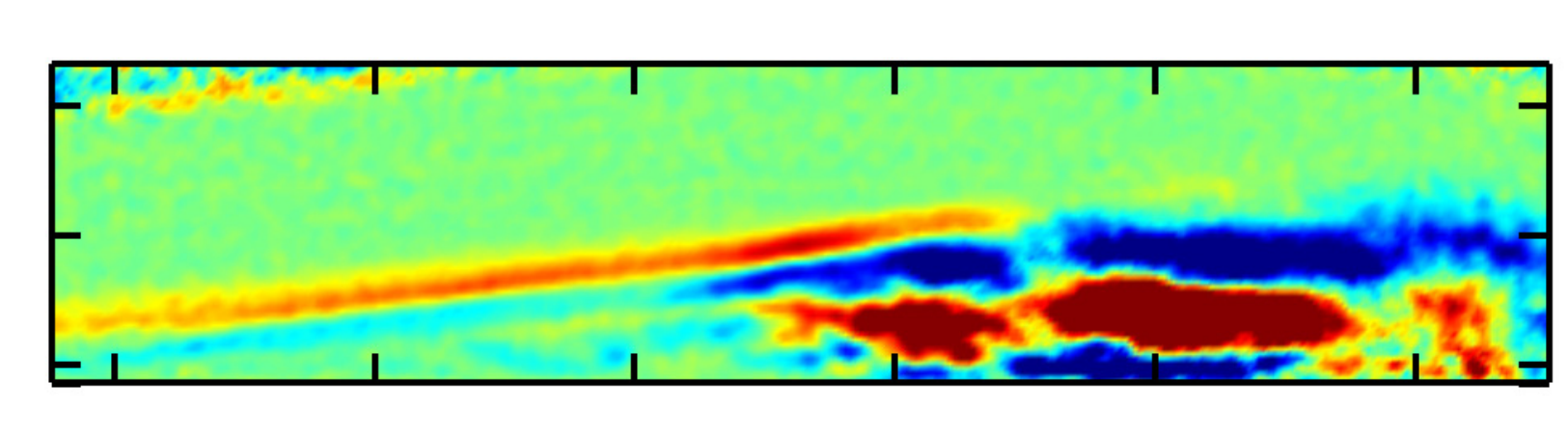}}
\end{center}
\end{minipage}
\caption{A comparison of TDMD modes with DMD modes suggests that some aspects of the separated flow dynamics may be characterized by different spatial structures than revealed by standard DMD;
while many TDMD modes are qualitatively similar to DMD modes, some of the modes differ.
The oscillatory vorticity modes are plotted top to bottom in order of decreasing frequency for DMD (a)--(i) and TDMD (j)--(r).
Here, $r=25$ and modes of vorticity are computed from DMD/TDMD modes of velocity.}
\label{fig:pivmodes}
\end{figure*}

\section{Concluding Discussion}
By representing DMD as a two-stage process, we have identified an asymmetric treatment of snapshot data in standard formulations of DMD.
As a result, we have isolated the source of noise-induced error in DMD that has previously been observed and reported in the literature.
Importantly, our determination of this error as a systematically introduced bias indicates that commonly employed approaches to ``de-noising,'' while reducing the variance in the resulting DMD analysis, will inevitably yield biased results;
the systematic introduction of bias errors cannot be removed by various methods for averaging and cross-validation.
Instead, we propose forming an augmented snapshot matrix~\eqref{eq:2}---as in problems of total least-squares---in order to account for the errors present in \emph{all} of the available data during the subspace projection step;
in doing so, one removes the systematic introduction of error and arrives at an unbiased formulation of DMD.
While the formulation proposed here is unbiased, further study is needed to robustify computational
analysis techniques; total least-squares problems are known to exhibit numerical sensitivity
due to their ``de-regularizing'' nature~\cite{fierroSIAMJSC1997,golubJNA1980,vanhuffelAutomatica1989,vanhuffel1991}.
The de-biasing procedure presented generalizes to other \emph{DMD-like}
algorithms as well; one need only apply the subspace projection step, 
and then replace the ``operator identification'' step by the
algorithm of choice (e.g., optimal mode decomposition~\cite{goulartCDC2012,wynnJFM2013}, streaming DMD~\cite{hematiPOF2014}, sparsity-promoting DMD~\cite{jovanovicPOF2014}, or optimized DMD~\cite{chenJNS2012}).
Despite these gains, techniques for distinguishing between measurement and process noise 
contributions to the analysis are still needed.

The availability of an unbiased DMD framework will be essential to performing valid data-driven Koopman spectral analysis in practical real-world contexts with imperfect snapshot measurements.
By invoking the two-stage TDMD framework, Koopman operator descriptions of a dynamical system determined from experimental data can be regarded with greater confidence, which will ultimately enable more accurate dynamical descriptions of complex time-evolving systems.
Moreover, forecasts of future system behavior from TDMD models will be more representative than those based on standard DMD models, since TDMD models will be able to ascertain the correct trends from past data, even when the data are noisy or imprecise.


\bibliographystyle{ieeetr}      
\bibliography{tlsdmd}   

\begin{thebibliography}{10}

\bibitem{schmidJFM2010}
P.~Schmid, ``Dynamic mode decomposition of numerical and experimental data,''
  {\em Journal of Fluid Mechanics}, vol.~656, pp.~5--28, 2010.

\bibitem{schmidAPS2008}
P.~Schmid and J.~Sesterhenn, ``Dynamic mode decomposition of numerical and
  experimental data,'' {\em 61st Annual Meeting of the {APS} {D}ivision of
  {F}luid {D}ynamics}, 2008.

\bibitem{rowleyJFM2009}
C.~W. Rowley, I.~Mezi\'c, S.~Bagheri, P.~Schlatter, and D.~Henningson,
  ``Spectral analysis of nonlinear flows,'' {\em Journal of Fluid Mechanics},
  vol.~641, pp.~115--127, 2009.

\bibitem{koopmanPNAS1931}
B.~O. Koopman, ``Hamiltonian systems and transformation in {H}ilbert space,''
  {\em Proceedings of the National Academy of Sciences}, vol.~17, no.~5,
  pp.~315--318, 1931.

\bibitem{koopmanPNAS1932}
B.~O. Koopman and J.~{von N}eumann, ``Dynamical systems of continuous
  spectra,'' {\em Proceedings of the National Academy of Sciences}, vol.~18,
  no.~3, pp.~255--263, 1932.

\bibitem{Neumann:1932a}
J.~von Neumann, ``Proof of the quasi-ergodic hypothesis,'' {\em Proceedings of
  the National Academy of Sciences}, vol.~18, pp.~70--82, 1932.

\bibitem{proctorIH2015}
J.~L. Proctor and P.~A. Eckhoff, ``Discovering dynamic patterns from infectious
  disease data using dynamic mode decomposition,'' {\em International Health},
  vol.~7, no.~2, pp.~139--145, 2015.

\bibitem{bourantasMP2014}
G.~C. Bourantas, M.~Ghommem, G.~C. Kagadis, K.~Katsanos, V.~C. Loukopoulos,
  V.~N. Burganos, and G.~C. Nikiforidis, ``Real-time tumor ablation simulation
  based on dynamic mode decomposition method,'' {\em Medical Physics}, vol.~41,
  no.~053301, 2014.

\bibitem{bruntonARXIV2014}
B.~W. Brunton, L.~A. Johnson, J.~G. Ojemann, and J.~N. Kutz, ``Extracting
  spatial-temporal coherent patterns in large-scale neural recording using
  dynamic mode decomposition,'' {\em ArXiv e-prints}, no.~1409.5496v1, 2014.

\bibitem{budisicCHAOS2012}
M.~Budi\u{s}i\'c, R.~Mohr, and I.~Mezi\'c, ``Applied {K}oopmanism,'' {\em
  Chaos}, vol.~22, no.~047510, 2012.

\bibitem{bergerAR2015}
E.~Berger, M.~Sastuba, D.~Vogt, B.~Jung, and H.~{Ben A}mor, ``Estimation of
  perturbations in robotic behavior using dynamic mode decomposition,'' {\em
  Advanced Robotics}, vol.~29, no.~5, pp.~331--343, 2015.

\bibitem{grosekARXIV2014}
J.~Grosek and J.~N. Kutz, ``Dynamic mode decomposition for real-time
  background/foreground separation in video,'' {\em ArXiv e-prints},
  vol.~1404.7592v1, 2014.

\bibitem{dukeEIF2012}
D.~Duke, J.~Soria, and D.~Honnery, ``An error analysis of the dynamic mode
  decomposition,'' {\em Experiments in Fluids}, vol.~52, no.~2, pp.~529--542,
  2012.

\bibitem{dukeJFM2012}
D.~Duke, D.~Honnery, and J.~Soria, ``Experimental investigation of nonlinear
  instabilities in annular liquid sheets,'' {\em Journal of Fluid Mechanics},
  vol.~691, pp.~594--604, 2012.

\bibitem{hematiPOF2014}
M.~S. Hemati, M.~O. Williams, and C.~W. Rowley, ``Dynamic mode decomposition
  for large and streaming datasets,'' {\em Physics of Fluids}, vol.~26,
  no.~111701, 2014.

\bibitem{schmidEIF2011}
P.~Schmid, ``Application of the dynamic mode decomposition to experimental
  data,'' {\em Experiments in Fluids}, vol.~50, pp.~1123--1130, 2011.

\bibitem{schmidTCFD2011}
P.~Schmid, L.~Li, M.~Juniper, and O.~Pust, ``Applications of the dynamic mode
  decomposition,'' {\em Theoretical and Computational Fluid Dynamics}, vol.~25,
  pp.~249--259, 2011.

\bibitem{semeraroEIF2012}
O.~Semeraro, G.~Bellani, and F.~Lundell, ``Analysis of time-resolved {PIV}
  masurements of a confined turbulent jet using {POD} and {Koopman} modes,''
  {\em Experiments in Fluids}, vol.~53, pp.~1203--1220, 2012.

\bibitem{tuJCD2014}
J.~H. Tu, C.~W. Rowley, D.~M. Luchtenburg, S.~L. Brunton, and J.~N. Kutz, ``On
  dynamic mode decomposition: Theory and applications,'' {\em Journal of
  Computational Dynamics}, vol.~1, pp.~391--421, December 2014.

\bibitem{wynnJFM2013}
A.~Wynn, D.~Pearson, B.~Ganapathisubramani, and P.~Goulart, ``Optimal mode
  decomposition for unsteady flows,'' {\em Journal of Fluid Mechanics},
  vol.~733, pp.~473--503, 2013.

\bibitem{fierroNLAA1995}
R.~D. Fierro and J.~R. Bunch, ``Orthogonal projection and total least
  squares,'' {\em Numerical Linear Algebra with Applications}, vol.~2, no.~2,
  pp.~135--153, 1995.

\bibitem{fierroLAIA1996}
R.~D. Fierro and J.~R. Bunch, ``Perturbation theory and orthogonal projection
  methods with applications to least squares and total least squares,'' {\em
  Linear Algebra and Its Applications}, vol.~234, pp.~71--96, 1996.

\bibitem{fierroSIAMJSC1997}
R.~D. Fierro, G.~H. Golub, P.~C. Hansen, and P.~{O'L}eary, ``Regularization by
  truncated total least squares,'' {\em {SIAM} Journal of Scientific
  Computing}, vol.~18, pp.~1223--1241, July 1997.

\bibitem{golubJNA1980}
G.~H. Golub and C.~F. {Van Loan}, ``An analysis of the total least squares
  problem,'' {\em SIAM Journal on Numerical Analysis}, vol.~17, pp.~883--893,
  December 1980.

\bibitem{golub1996}
G.~H. Golub and C.~F. {Van Loan}, {\em Matrix Computations}.
\newblock Baltimore, MD: Johns Hopkins University Press, 3~ed., 1996.

\bibitem{markovskySP2007}
I.~Markovsky and S.~{Van Huffel}, ``Overview of total least squares methods,''
  {\em Signal Processing}, vol.~87, pp.~2283--2302, October 2007.

\bibitem{vanhuffel1991}
S.~{Van Huffel} and J.~Vandewalle, {\em The Total Least Squares Problem:
  Computational Aspects and Analysis}.
\newblock Frontiers in Applied Mathematics, 9, Philadelphia, PA: SIAM, 1991.

\bibitem{zoltowskiSPIE1988}
M.~D. Zoltowski, ``Generalized minimum norm and constrained total least squares
  with applications to array signal processing,'' {\em Proceedings of SPIE},
  vol.~975, pp.~78--85, 1988.

\bibitem{goulartCDC2012}
P.~Goulart, A.~Wynn, and D.~Pearson, ``Optimal mode decomposition for high
  dimensional systems,'' in {\em 51st IEEE Conference on Descision and
  Control}, 2012.

\bibitem{jovanovicPOF2014}
M.~R. Jovanovi\'c, P.~J. Schmid, and J.~W. Nichols, ``Sparsity promoting
  dynamic mode decomposition,'' {\em Physics of Fluids}, vol.~26, no.~024103,
  2014.

\bibitem{chenJNS2012}
K.~Chen, J.~Tu, and C.~Rowley, ``Variants of dynamic mode decomposition:
  Boundary condition, {K}oopman, and {F}ourier analysis,'' {\em Journal of
  Nonlinear Science}, vol.~22, no.~6, pp.~887--915, 2012.

\bibitem{bagheriPOF2014}
S.~Bagheri, ``Effects of weak noise on oscillating flows: Linking quality
  factor, {F}loquet modes, and {K}oopman spectrum,'' {\em Physics of Fluids},
  vol.~26, no.~094104, 2014.

\bibitem{mezicND2005}
I.~Mezi\'c, ``Spectral properties of dynamical systems, model reduction and
  decompositions,'' {\em Nonlinear Dynamics}, vol.~41, no.~1--3, pp.~309--325,
  2005.

\bibitem{mezicAnnRev2013}
I.~Mezi\'c, ``Analysis of fluid flows via spectral properties of the {K}oopman
  operator,'' {\em Annual Review of Fluid Mechanics}, vol.~45, pp.~357--378,
  2013.

\bibitem{williamsAIMS2014}
M.~O. {Williams}, I.~G. {Kevrekidis}, and C.~W. {Rowley}, ``{A Data-Driven
  Approximation of the Koopman Operator: Extending Dynamic Mode
  Decomposition},'' {\em ArXiv e-prints}, Aug. 2014.

\bibitem{williamsPR2014}
M.~O. {Williams}, C.~W. {Rowley}, and I.~G. {Kevrekidis}, ``A kernel-based
  approach to data-driven {K}oopman spectral analysis,'' {\em ArXiv e-prints},
  November 2014.

\bibitem{gleserAS1981}
L.~J. Gleser, ``Estimation in a multivariate ``error-in-variables'' regression
  model: Large sample results,'' {\em The Annals of Statistics}, vol.~9, no.~1,
  pp.~24--44, 1981.

\bibitem{vanhuffelAutomatica1989}
S.~{Van Huffel} and J.~Vandewalle, ``On the accuracy of total least squares and
  least squares techniques in the presence of errors on all data,'' {\em
  Automatica}, vol.~25, no.~5, pp.~765--769, 1989.

\bibitem{griffinThesis2013}
J.~C. Griffin, {\em On the Control of a Canonical Separated Flow}.
\newblock PhD thesis, University of Florida, 2013.

\bibitem{griffinAIAA2013}
J.~Griffin, M.~Oyarzun, L.~N. Cattafesta, J.~H. Tu, C.~W. Rowley, and
  R.~Mittal, ``Control of a canonical separated flow,'' in {\em AIAA Paper
  2013-2968}, 2013.

\end{thebibliography}


\end{document}